\documentclass[preprint]{aastex}

\usepackage{graphicx}

  \makeatletter
  \def\mathcomposite{%
     \@ifstar
        {\def\@mathcomposite@option{%
            \baselineskip\z@skip\lineskiplimit-\maxdimen}%
         \@mathcomposite}%
        {\let\@mathcomposite@option\offinterlineskip
         \@mathcomposite}}
  \def\@mathcomposite{%
     \@ifnextchar[\@@mathcomposite{\@@mathcomposite[0]}}
  \def\@@mathcomposite[#1]#2#3#4{%
     #2{\mathchoice
        {\@mathcomposite@{#1}{#3}{#4}\displaystyle{1}}%
        {\@mathcomposite@{#1}{#3}{#4}\textstyle{1}}%
        {\@mathcomposite@{#1}{#3}{#4}%
         \scriptstyle\defaultscriptratio}%
        {\@mathcomposite@{#1}{#3}{#4}%
         \scriptscriptstyle\defaultscriptscriptratio}}}
  \def\@mathcomposite@#1#2#3#4#5{%
     \vcenter{\m@th\@mathcomposite@option
        \dimen@\f@size\p@\dimen@#1\dimen@\dimen@#5\dimen@
        \divide\dimen@ 18
        \edef\@mathcomposite@skipamount{\the\dimen@}%
        \ialign{\hfil$#4##$\hfil\cr
           #2\crcr
           \noalign{\vskip\@mathcomposite@skipamount}%
           #3\crcr}}}
  \makeatother

\shorttitle{Formation of SiC Grains in C--rich AGB Stars}
\shortauthors{Yasuda \& Kozasa}

\begin{document}


\title{Formation of SiC Grains  in Pulsation--Enhanced Dust--Driven Wind \\
    around Carbon--Rich Asymptotic Giant Branch Stars}


\author{Yuki Yasuda and Takashi Kozasa}
\affil{Department of Cosmosciences, Graduate School of Science, Hokkaido University, Sapporo 060--0810, Japan}
\email{yuki@antares-a.sci.hokudai.ac.jp}

%
%
%

%

\begin{abstract}
We investigate the formation of silicon carbide (SiC) grains in the framework of dust--driven wind
around pulsating carbon--rich Asymptotic Giant Branch (C--rich AGB) stars 
in order to reveal not only the amount but also the size distribution.
Two cases are considered for the nucleation process;
one is the LTE case where the vibration temperature of SiC clusters $T_{\rm v}$
is equal to the gas temperature as usual, and another is the non--LTE
case in which $T_{\rm v}$ is assumed to be the same as the temperature of small SiC grains.
The results of hydrodynamical calculations for a model with stellar parameters of 
mass $M_{\ast}$=1.0 $M_{\odot}$, luminosity $L_{\ast}$=10$^{4}$ $L_{\odot}$,
effective temperature $T_{\rm eff}$=2600 K, C/O ratio=1.4,
and pulsation period $P$=650 days
show the followings: 
In the LTE case, SiC grains condense in accelerated outflowing gas after
the formation of carbon grains and the resulting averaged mass ratio of SiC to
carbon grains of $\sim$ 10$^{-8}$ is too small to reproduce the value of
0.01--0.3 inferred from the radiative transfer models. 
On the other hand, in the non--LTE case, the formation region of SiC grains
is inner than and/or almost identical to that of carbon grains due to 
the so--called inverse greenhouse effect.
The mass ratio of SiC to carbon grains averaged at the outer boundary
ranges from 0.098 to 0.23 for the sticking probability $\alpha_{\rm s}$=0.1--1.0. 
The size distributions with the peak at $\sim$ 0.2--0.3 $\rm{\mu}$m in radius
cover the range of size derived from the analysis of presolar SiC grains.
Thus the difference between temperatures of small cluster and gas plays a crucial
role in the formation process of SiC grains around C--rich AGB stars, and
this aspect should be explored for the formation process of dust
grains in astrophysical environments.
\end{abstract}





\section{INTRODUCTION}

Presolar silicon carbide (SiC) is the abundant and well 
studied dust species among presolar
grains extracted from meteorites. Based on the 
isotopic compositins of C, N, Si and Al, presolar SiC grains are
classified into six categories named ``mainstream'', A, B, X, Y and Z
grains (e.g., see \citealt{ott10} for a review). Among them, the 
mainstream grains populating 
about 90 \% of presolar SiC grains have been considered to be 
originated in 
carbon--rich Asymptotic Giant Branch (C--rich AGB) stars from  
the isotopic compositions
of C, N,  and Si \citep{zin87,hop94} and 
the s--process signatures in the noble
gases as well as in the heavy elements \citep{hop97}.
\citet{ama94} have reported that SiC grains extracted 
from the Murchison meteorite have a log--normal distribution 
with the peak at $\sim$ 0.4 $\micron$ in diameter,  and 70 \% 
of them by  number are in the grains whose diameter 
ranges from 0.3 to
0.7 $\micron$ \citep{dau03}. 
Although it was reported that SiC grains are  
more fine grained in other meteorites than in Murchison 
(see \citealt{rus97}), a significant 
fraction of presolar SiC grains are larger than 0.3 
$\micron$ in diameter (see \citealt{hop00}). Thus, submicron--sized 
SiC grains are expected to form around C--rich AGB stars. 
The presence of SiC grains in C--rich AGB stars was 
confirmed from
the emission feature around 11.3 $\micron$ \citep{hac72,tre74,goe80}, 
prior to the discovery of presolar SiC grains \citep{ber87}. 
Also, the absorption feature attributed to SiC grains
was observed in several extreme carbon stars \citep{jon78,spe97,spe09}.
The radiative transfer models fitting the observed spectral energy 
distributions (SEDs) including  the spectral feature attributed to SiC grains 
have estimated the mass ratio of  SiC to carbon grains
to be in the range of 0.01--0.3
\citep{lor93,lor94,bla94, bla98, gro95,gro98}.  
The variation of the mass ratio of 
SiC to carbon derived from the radiative transfer models arises from 
the grain models used, and the value could be also 
influenced by the optical constants of SiC used in the radiative
transfer models (see the discussion in Section 6).

Although the isotopic signatures of presolar SiC grains and 
the observed spectral feature around 11.3 $\micron$
support the formation of SiC grains in circumstellar 
envelopes of C--rich AGB stars, the formation process as well  
as the formation conditions remain unknown and 
debatable: 
Thermodynamic equilibrium calculations have been applied
to investigate the formation condition of dust grains in 
C--rich AGB stars (e.g. \citealt{lod95}), and the calculations have
claimed that high gas pressure 
(0.3--300 dyne cm$^{-2}$) and low C/O ratio 
(1.0 $<$ C/O $\lesssim$ 1.1) is necessary for reproducing 
the condensation sequence of presolar TiC, graphite, 
and SiC grains in this order inferred from the presence
of graphitic spherules containing a TiC core \citep{sha95,dau03}. 
On the other hand, \citet{mcc82} suggested  
the formation of SiC grains earlier than carbon grains, 
considering the inverse greenhouse effect that the 
temperature of SiC grains which is  
transparent in the optical to near infrared (NIR) region 
\citep{spi59,hof09} is lower than the temperature of carbon grains in 
circumstellar envelopes of 
C--rich AGB stars. In addition, the interferometric observation of 
IRC+10216 at 11 $\micron$ has suggested the presence of 
dust species more transparent than carbon grains in the 
inner circumstellar envelope close to the photosphere \citep{dan90}. 
Although thermodynamic equilibrium calculations 
can tell us the physical conditions in which a condensate exists 
stably, the calculations as well as the astronomical
observations provide directly no information on the number 
and size of grains that are essential to 
clarify the physical conditions prevailing in the formation sites  
of grains and the physical and chemical processing 
suffering in the interstellar space. Also, it should be pointed out here
that the C/O ratio in C--rich AGB stars is not restricted to be less than 1.1, 
and tends to increase with  decreasing the effective temperature 
$T_{\rm eff}$; the low C/O ratio observed in stars with 
$T_{\rm eff}$ $\lesssim$ 2500 K 
is considered to be caused by the
neglect of formation of dust in the 
model atmosphere  \citep{ber02,ber05}.

So far dust formation around C--rich AGB stars has been investigated 
in the framework of pulsation--enhanced dust--driven wind models 
since the study by \citet{fle92}, and it has been believed  
that the hydrodynamical models including only the formation of carbon grains 
well reproduce the observed dynamical behavior of C--rich AGB 
stars whose mass loss rates exceed 
10$^{-6}$ $M_{\odot}$ yr$^{-1}$ 
(e.g., \citealt{win94,win97,now05}).  
However, the formation of SiC grains has not yet 
pursued 
in the hydrodynamical models. On the other hand, 
formation of a variety of dust species including SiC 
grains has been 
studied in the steady state dust--driven wind models around AGB 
stars in order to investigate the evolution of dust in the 
interstellar space (see \citealt{fer06,zhu08a}). 
However, in the models, the number of seed nuclei is 
assumed in the formation process. Although the models may predict 
the mass of dust species  supplied from AGB stars, 
no information is available for the size distribution. 
Taking into account the nucleation process and 
including the so--called inverse greenhouse effect, 
\citet{koz96} have investigated the formation of SiC grains 
in an outflowing gas with  constant velocity around C--rich 
AGB stars and shown that SiC grains condense prior to 
carbon grains for the mass--loss rate smaller than $\sim$ 
1.5 $\times$ 10$^{-5}$ $M_{\odot}$ yr$^{-1}$; otherwise 
composite grains consisting of carbon and SiC form. 
The calculated radius of $\sim$ 0.01 $\micron$ is 
too small to reproduce the size of presolar SiC grains 
and the proposed grains consisting of a SiC core and a 
carbon mantle have not been discovered 
except for one in presolar grains \citep{cro10}. 
Thus the investigation of formation of SiC grains in a more 
realistic hydrodynamical model remains an  
important subject to be explored which will reveal the formation 
process and condition around C--rich AGB stars. 
The knowledge of not only the amount but also the size distribution 
of dust grains is crucial for investigating the 
role of dust in the universe as shown by \citet{yam11}.  
In particular, the size distribution of SiC 
grains formed around C--rich AGB stars is vital 
to get insight into the evolution and processing 
during the journey from the formation sites to 
the incorporation into meteorites 
from the analysis of presolar SiC grains.

In this paper, we aim at exploring the formation process of 
SiC grains and investigating the amount as well as the size distribution 
in the framework of pulsation--enhanced dust--driven wind 
from C--rich AGB stars. 
We develop the hydrodynamical model including formation of carbon 
and SiC grains as well as  a non--grey radiative transfer. 
The formulation by \citet{gau90} is applied for 
the nucleation and growth processes of carbon 
grains\footnote{In the
formulation it is assumed that carbon grains nucleate and grow
homogeneously. The recent analysis of presolar graphites
\citep{cro05} reavealed that about 40 \% of the carbide--containing
graphites have a carbide core at their center. This implies
that at least more than 60 \% of presolar graphites nucleate and grow
homogeneously.}.
We assert that SiC grains  nucleate and grow 
homogeneously starting from SiC molecule because   
all presolar SiC grains whose internal structure has thus far been 
analyzed do not contain any seed nuclei at their centers
\citep{str05}. Furthermore there is no evidence 
in presolar graphite grains for a SiC seed nuclei 
except for one grain \citep{cro10}.
The two cases are considered for the formation process of 
SiC grains; one is the LTE case where the vibration 
temperature $T_{\rm v}$ of SiC cluster is the same as 
the gas temperature as usual. 
Another is the non--LTE case taking into  
account the inverse greenhouse effect on the 
formation, in which  $T_{\rm v}$ is assumed to be 
the same as the temperature of small SiC grains. We 
will show that the mass ratio of SiC to carbon grains and 
the radius of SiC grains are at most 10$^{-6}$ and 
0.1 $\micron$, respectively, in the LTE case, while 
the mass ratio ranges from 0.098 to 0.23 and the 
typical radius ranges from 0.2 to 0.6 $\micron$ for 
the sticking probability $\alpha_{\rm s}$ of 0.1--1.0 
in the non--LTE case. 

This paper is organized as follows; The chemical reactions and 
gas species relevant to the formation of SiC grains are examined 
and the formation process of SiC grains is formulated in Section 2, 
and the numerical scheme for calculation of grain formation 
is presented in Section 3, 
including the method for tracing the size distribution.
Section 4 describes the hydrodynamical model and 
the modeling procedure for pulsation--enhanced dust--driven 
wind model including dust formation and radiative transfer.  
The results of calculations are presented in Section 5 
and are discussed in Section 6. Summary is presented in Section 7.

\section{FORMATION PROCESS OF SIC GRAINS}

In the hydrodynamical models of
pulsation--enhanced dust--driven wind from C--rich
AGB stars so far developed since the work of \citet{fle92},
the formation of carbon grains has been processed according to the 
recipe developed by \citet{gau90};
the abundances of gas species responsible for the nucleation and growth
of carbon grains are assumed to be in chemical
equilibrium in the gas whose elemental abundance is solar 
except for carbon.
In the recipe based on \citet{gai88},
the formation of clusters with number of constituent molecules (size)  
up to $n_{\rm out}$=1000 is evaluated from the steady--state 
nucleation rate, and the growth and 
evaporation of grains with size $n \ge n_{\rm out}$ is treated as 
macroscopic grains. The formation of carbon grains 
are considered to proceed through the two body reactions of 
C$_2$H$_2$ and C$_2$H molecules with carbon clusters,   
starting from atomic carbon or diatomic carbon molecule.

In this section, we shall explore the formation process 
of SiC grains around C--rich AGB stars, following the 
formulation for the formation of carbon grains \citep{gai88,gau90}.
First we examine the gas species responsible for the nucleation and 
the growth processes of SiC grains from chemical equilibrium 
calculations. In the same way as the formation of carbon grains,  
we consider that the nucleation process proceeds through the two body
reactions of cluster with gas species, starting from the molecules 
whose stoichiometric composition is the  same as SiC grains. 
On the other hand, we allow three or more body reactions on the 
surface for the growth process. Then, the basic 
equations describing the 
nucleation and growth processes of SiC grains 
are presented.

\subsection{Si--bearing gas species in dust formation region}

Figure 1 shows the temperature dependence of 
relative abundances of Si--bearing molecules (left)  and 
C--bearing molecules (right) in chemical equilibrium 
in the gas with the density $\rho=$ 10$^{-13}$ g cm$^{-3}$ 
and C/O = 1.4, 
which is calculated by the method of the minimization of the 
Gibbs free energies. The elemental abundances except for carbon are 
taken from \citet{all73}, and the thermodynamic data are taken 
from 
\citet{cha85}
except for C$_2$H$_2$ \citep{che92}, SiC$_{3}$, SiC$_{4}$, SiC$_{5}$, Si$_{2}$C$_{2}$, Si$_{2}$C$_{3}$,
Si$_{2}$C$_{4}$, Si$_{3}$C, Si$_{3}$C$_{2}$, Si$_{3}$C$_{3}$, Si$_{4}$C,
Si$_{4}$C$_{2}$, and Si$_{5}$C \citep{den08}.
We can see 
that,  
as the starting molecules for the formation of SiC grains,
SiC is most abundant in the region whose gas temperature is higher than 1350 K, 
while Si$_{2}$C$_{2}$ molecule is abundant in the lower temperature region.
Among Si--bearing molecules that are considered to be 
candidate reactants in the growth process of SiC grains,
Si, SiC$_{2}$, and SiS molecules are abundant 
in the overall 
range of gas temperature presented in Figure 1, and 
Si$_{2}$C is also abundant in the low temperature region of $T<$ 1100 K. 
As shown in the right panel, 
hydrocarbon molecules such as C$_{2}$H$_{2}$ or C$_{2}$H are expected to be
reactants participating in the growth process 
on the surface of SiC grains unless these molecules are 
heavily depleted by the formation of carbon grains.

\subsection{Nucleation process of SiC grains}

Including the starting molecules SiC and Si$_{2}$C$_{2}$ 
for the nucleation process, 
the candidate reactants for the two body reaction
in the nucleation process 
are confined to twelve Si--bearing molecules: 
SiC, SiC$_{2}$, SiC$_{3}$,
SiC$_{4}$, Si$_{2}$C, Si$_{2}$C$_{2}$, Si$_{2}$C$_{3}$, Si$_{3}$C,
Si$_{3}$C$_{2}$, Si$_{4}$C, Si$_{4}$C$_{2}$, and Si$_{5}$C.
We shall exclude the gas species with which the reaction 
for formation of SiC clusters is endothermic.
Table 1 presents the reaction enthalpies $\Delta H{\arcdeg}$ 
of the twelve candidates 
for the formation of Si$_{2}$C$_{2}$ and Si$_{3}$C$_{3}$ at 1000 and 1500 K.
The molecules SiC$_{2}$, SiC$_{3}$, SiC$_{4}$, Si$_{2}$C, Si$_{2}$C$_{3}$,
Si$_{3}$C, Si$_{4}$C, and Si$_{5}$C are excluded from the candidates 
since the reactions are endothermic.
The reaction enthalpies for the formation of SiC
clusters with $n>3$ for which no thermodynamic data is available 
are replaced with those for the formation of solid SiC, 
and the values at 1000  and 1500 K are listed in Table 2  
where the thermodynamic data  of $\beta$--SiC\footnote{
Presolar SiC grains so far analyzed show only two polytypic forms 
\citep{dau03}: $\beta$--SiC (79.4 \% by number), $\alpha$--SiC (2.7 \%), 
and the intergrowths of these two forms (17.1 \%). Note that the 
difference of thermodynamic data between $\alpha$--SiC and  $\beta$--SiC
does not affect the results of calculations presented in this paper.
} is used in the 
calculations. The molecules SiC$_{2}$ and  Si$_{5}$C are excluded 
because of the endothermic reactions.
Among the remaining candidates, as the reactants in the nucleation 
process of SiC grains,  we adopt SiC and Si$_{2}$C$_{2}$
since these molecules are much more abundant than the others in the 
temperature region of interest to us as shown 
in Figure 1 (left).

We discard the nucleation process starting from the two 
body reaction between Si$_{2}$C$_{2}$ molecules 
since Si$_{2}$C$_{2}$ is relatively stable and without the 
thermodynamic data of Si$_{4}$C$_{4}$ it is difficult to judge 
whether the reaction for the formation of Si$_{4}$C$_{4}$ 
proceeds or not.
Thus we consider the following two reactions in the nucleation process of
SiC grains:
\begin{eqnarray}
\mathrm{(SiC)_{\it n}}+\mathrm{SiC} \rightarrow \mathrm{(SiC)_{{\it n}+1}}\\
\mathrm{(SiC)_{2{\it n}-1}}+\mathrm{(SiC)_{2}} \rightarrow \mathrm{(SiC)_{2{\it n}+1}}
\end{eqnarray}
where $n \ge 1$, and note that we take SiC as the starting molecule for 
the nucleation process of SiC grains. 
We shall discuss the effect of 
the nucleation process starting from the 
two body reaction of Si$_{2}$C$_{2}$ molecules  in Section 6.

In the nucleation process, clusters grow 
through the attachment of reactant and the 
growth rate depends on the number density and kinetic temperature 
of the reactants.
On the other hand, clusters decay through the detachment of 
reactant molecules, and  
the decay rate depends on the vibration states of 
the cluster \citep{gai88}.
As pointed out by \citet{koz96}, 
in the dust formation region around C--rich AGB stars,
the vibration temperature of SiC
cluster $T_{\rm v}$ is not always the same as the kinetic temperature of
gas $T$ due to the low gas density and strong radiation field.
The principle of detailed balance being applied,
the decay rate can be related with the growth rate
by considering a hypothetical situation in which the cluster is in a local thermal
equilibrium with the ambient gas whose temperature is $T_{\rm v}$.
Then, under the assumption that 
$T_{\rm v}$ is independent of the size of cluster,  
the steady--state nucleation rate $J^{i}_{{\rm S}}$ 
for the reaction with (SiC)$_{\it i}$ is given by
\begin{eqnarray}
J^{i}_{{\rm S}}=4\pi\alpha_{{\rm s}}^{i}
a_{0}^{2}n_{\rm SiC}n_i\sqrt{\frac{k_{{\rm B}}T}{2\pi m_{i}}}
\times\left[1+\displaystyle\sum^{N_{{\rm x}}}_{k=1}
\frac{\exp\biggl\{4\pi
a_{0}^{2}(ik)^{2/3}\sigma_{ik+1}/k_{{\rm B}}T_{\rm v}\biggr\}}
{\left(ik+1\right)^{2/3}
\biggl\{S_{i}(T_{\rm v},P_{i})\sqrt{T/T_{\rm v}}\biggr\}^{k}}\right]^{-1}
\end{eqnarray}
where $\it i$=1 (2) denotes the reaction (1) (the reaction (2)) 
and the quantities related with the reactant SiC (Si$_{2}$C$_{2}$) molecule,
$a_{0}$ is the hypothetical radius of a monomer in the bulk phase,
$n_{\rm SiC}$ the number density of starting molecule SiC. 
The number density, sticking probability, mass, and 
partial pressure of the $\it i$--th reactant is represented by 
$n_i$, $\alpha_{\rm s}^{i}$, $m_{i}$, and $P_i$, respectively.  
The Boltzmann constant is $k_{\rm B}$, 
$N_{\rm x}=\lfloor(n_{{\rm out}}-2)/i\rfloor$ 
with the floor function $\lfloor x\rfloor$, 
and the surface tension of the cluster of size $\it n$ $\sigma_{n}$. 
The supersaturation ratio 
$S_{i}(T_{\rm v},P_i)$ is given by
\begin{eqnarray}
\ln S_{i}(T_{\rm v},P_i)=-\frac{\triangle G^{0}_{i}}{k_{\rm B}T_{\rm v}}+\ln\frac{P_i}{P_{\rm 0}}.
\end{eqnarray}
where  $P_{\rm 0}$ is the standard pressure.
In the above equation,
$\triangle G^{0}_{i}$ is the Gibbs free energy for the formation of bulk SiC
from the ${\it i}$--th reactant in the standard pressure,  
and the temperature dependence of 
$\triangle G^{0}_{i}/k_{\rm B}T_{\rm v}$ is evaluated from the least--squares
fitting given by 
\begin{eqnarray}
\frac{\triangle G^{0}_{i}}{k_{\rm B}T_{\rm v}}=\alpha_{i}
+\beta_{i}\ln T_{\rm v}
+\frac{\gamma_{i}}{T_{\rm v}^{0.5}}
+\frac{\delta_{i}}{T_{\rm v}}+\frac{\eta_{i}}{T_{\rm v}^{2}}
+\frac{\epsilon_{i}}{T_{\rm v}^{3}}.
\end{eqnarray}
whose form is based on a polynomial expression of heat capacity 
at the constant pressure \citep{ber85}.
The coefficients
$\alpha_{i}$, $\beta_{i}$, $\gamma_{i}$, $\delta_{i}$, $\eta_{i}$ and $\epsilon_{i}$
are given in Table 3.
In the calculations, we adopt $a_{0}$=1.71$\times$10$^{-8}$ cm which is
evaluated from the bulk density of solid SiC ($\rho$=3.16 g cm$^{-3}$) 
taken from  the CRC hand book \citep{lid03}.
The values of the sticking probabilities of SiC and 
Si$_{2}$C$_{2}$ molecules on 
the SiC cluster are unknown and are  assumed to be unity. 
The surface tension $\sigma_{n}$ of SiC cluster with size 
${\it n}$=2 or 3 is directly evaluated 
from the following definition of surface tension of $\it n$--mer
\begin{eqnarray}
4\pi a_{0}^{2}(n-1)^{\frac{2}{3}}\sigma_{n}=\mathring{g}(n)-\mathring{g}(1)-(n-1)\mathring{g}_{s}
\end{eqnarray}
where $\mathring{g}(n)$ and $\mathring{g}_{s}$ 
are the Gibbs free energies of $\it n$--mer in gas phase and the monomer 
in bulk phase in the standard state, respectively.
On the other hand, we adopt the bulk value of $\sigma$=840 erg cm$^{-2}$
measured by \citet{all59} as the surface tension of SiC clusters with 
size ${\it n}>3$.

\subsection{Growth of SiC grains}

We allow three or more body reactions for the growth of SiC grains   
in addition to the two body reactions for the nucleation. 
As shown in Figure 1, the abundant Si-- and C--bearing molecules 
expected to be the reactants for the grain growth are  
Si, SiC$_{2}$, Si$_{2}$C, C, C$_{2}$H$_{2}$, and C$_{2}$H, 
except for CO and SiS which are stable against the chemical reactions.
Then we select the following exothermic reactions;
\begin{eqnarray}
\mathrm{(SiC)_{\it n}}+\mathrm{SiC} \rightarrow \mathrm{(SiC)_{{\it n}+1}},\\
\mathrm{(SiC)_{\it n}}+\mathrm{Si_{2}C_{2}} \rightarrow \mathrm{(SiC)_{{\it n}+2}},\\
\mathrm{(SiC)_{{\it n}}+2}\mathrm{Si}+\mathrm{C_{2}H_{2}} \rightarrow \mathrm{(SiC)_{{\it n}+2}}+\mathrm{H_{2}},\\
\mathrm{(SiC)_{{\it n}}+2}\mathrm{Si}+\mathrm{C_{2}H}+\mathrm{H} \rightarrow \mathrm{(SiC)_{{\it n}+2}}+\mathrm{H_{2}},\\
\mathrm{(SiC)_{{\it n}}+2}\mathrm{Si_{2}C}+\mathrm{C_{2}H_{2}} \rightarrow \mathrm{(SiC)_{{\it n}+4}}+\mathrm{H_{2}},\\
\mathrm{(SiC)_{{\it n}}+}\mathrm{Si}+\mathrm{SiC_{2}} \rightarrow \mathrm{(SiC)_{{\it n}+2}},\\
\mathrm{(SiC)_{{\it n}}+}\mathrm{Si}+\mathrm{C} \rightarrow \mathrm{(SiC)_{{\it n}+1}}
\end{eqnarray}
where  $\it n>n_{\rm out}$=1000; 
the reaction enthalpies at 1000 K
are -1265, -1739, -1572, -1206, and -1240 kJ mol$^{-1}$
for the chemical reactions from (9) to (13) in this order.

Except for the two body reactions, it is assumed that 
each of the chemical reactions proceeds through the attachment 
of a key species which is defined as 
the species of the least collision frequencies among the reactants and
is considered to control the kinetics of growth process 
\citep{koz87,has88}.
Then, the time evolution of the size of dust grains with radius $a$ 
is given by 
\begin{eqnarray}
\frac{d}{dt}\left(\frac{a}{a_{0}}\right)=\frac{1}{3\tau_{{\rm net}}}
\end{eqnarray}
where the size--independent net time scale of grain 
growth $\tau_{{\rm net}}$ \citep{gau90} 
is the sum of the net time scales for the reactions from (7) to (13) 
and is written as follows: 
\begin{eqnarray}
\frac{1}{\tau_{{\rm net}}}=\displaystyle\sum_{i=1}^{2}\frac{1}{\tau^{i}_{{\rm net}}}
+\displaystyle\sum_{n=9}^{13}\frac{1}{\tau^{{\rm C},n}_{{\rm net}}}.
\end{eqnarray}
In the above equation, $\tau^{i}_{{\rm net}}$ (${\it i}$=1,2) 
is the net time scale of grain growth by the two body 
reaction with (SiC)$_{\it i}$ molecule and given by 
\begin{eqnarray}
\frac{1}{\tau^{i}_{{\rm net}}}=i4\pi a^{2}_{0}\alpha_{{\rm s}}^{i}
\sqrt{\frac{k_{{\rm B}}T}{2\pi m_{i}}}
\left(n_i-\frac{P_{{\rm ev}}}{k_{{\rm B}}\sqrt{TT_v}}\right),
\end{eqnarray}
and $\tau^{{\rm C},n}_{{\rm net}}$ (n=9 to 13) is the net time scale of
grain growth for each of the chemical reactions from (9) to (13) 
and is given in the same form
of equation (16) by replacing $\alpha_{{\rm s}}^{i}$, $m_i$ and $n_{i}$
with those for the  key species.
In equation (16), $P_{{\rm ev}}$ represents 
the vapor pressure for the key species in the reaction; 
for  example, in the reactions (9) and (10),
the vapor pressure of Si atom is assumed to be 
the same as that for the reaction: 
$\mathrm{(SiC)_{\rm s}} \rightarrow \mathrm{Si}+\mathrm{C}$.
In the reaction (11), the vapor pressure of Si$_{2}$C 
molecule is assumed to be the same as that 
for the reaction: 
$2\mathrm{(SiC)_{\rm s}} \rightarrow \mathrm{Si_{2}C}+\mathrm{C}$.
In the calculations, the sticking probability 
$\alpha_{{\rm s}}$ is assumed to be unity
for all reactions.

\section{NUMERICAL SCHEME FOR GRAIN FORMATION}

The  number densities of the gas species responsible for the 
formation process of carbon and SiC grains in the pulsation--enhanced
dust--driven wind from C--rich AGB stars are evaluated from the chemical
equilibrium calculations for given gas density and temperature coupled with 
the hydrodynamical calculation. 
We include 4 atoms (H, C, Si, S) and
14 molecules (H$_{2}$, C$_{2}$, C$_{3}$, C$_{2}$H, C$_{2}$H$_{2}$, SiH,
SH, H$_{2}$S, SiC, SiC$_{2}$, Si$_{2}$C, Si$_{2}$C$_{2}$, CS, SiS) in
the chemical equilibrium calculations, and 
the abundances of elements except for carbon are 
taken from the table of \citet{all73}.
We employ the formulation presented in Section 2 for the formation 
process of SiC grains, 
while the formulation by \citet{gau90} is applied for the formation of
carbon grains. As mentioned before, the sticking probability 
$\alpha_{\rm s}$ for all reactions at nucleation and growth of 
SiC grains is assumed to be unity in the calculations, whose 
effect on the results of calculations is discussed in Section 6.

The number density and size of grains as well as the fraction of
condensible molecules locked into grains are evaluated by taking 
the $j$--th moment of the size distribution of grains  per H--element 
$\hat{K}_{j}$ defined as 
\begin{eqnarray}
\hat{K}_{j}=\displaystyle\sum^{\infty}_{n=n_{{\rm out}}}N^{\frac{j}{3}}\frac{f(n,t)}{n_{<{\rm H}>}}
\end{eqnarray}
where $f(n,t)$ is the number density of grains with the size $n$ at time $t$
and $n_{<{\rm H}>}$=$n_{\rm H}$+2$n_{\rm H_{2}}$. 
Normally, the moments are calculated by reducing to the simultaneous 
differential equations \citep{gai88}. 
The time variation of gas (grain) temperature is not always 
monotonic and rather complicated in the  
pulsation--enhanced dust--driven winds.
In the case that the formed grains are being lost by 
the destruction due to evaporation at time $t$,
the extinction term is included in the differential equations 
by using 
the number density of grains 
with the minimum size $f(n_{\rm out},t)$ 
which has to be derived by tracing the size distribution.
Therefore, solving the differential equations must be compensated with 
the evaluation of the size distribution
of grains with a limited number of size bins
in the hydrodynamical simulations \citep{fle94}.

Thus, in this paper, we calculate the moments 
by employing the finite and adaptive size grids rather than solving a set of 
the simultaneous differential equations. 
Each of the size bins is comoved in size space \citep{kr95,woi05}.
The $i$--th 
size bin is characterized by 
the number of grains per H--element $\hat{w}(i)$ 
in the size interval with lower limit $a_{i,{\rm min}}$ 
and upper limit $a_{i,{\rm max}}$. 
The lower and upper limits are calculated by integrating equation (14).
When the net time scale $\tau_{{\rm net}}$ is positive,
the number of  grains per H--element in the lowest size bin $\hat{w}(I)$ where $I$
is the total number of size bins is 
given by integrating the equation 
\begin{eqnarray}
\frac{d\hat{w}(I)}{dt}=\frac{J_{{\rm S}}}{n_{<{\rm H}>}}.
\end{eqnarray}
When the net time scale $\tau_{{\rm net}}$ is negative, with the destruction rate of 
grains $J_{\rm des}$ estimated by
\begin{eqnarray}
J_{{\rm des}}=-\frac{da}{dt}\frac{\hat{w}(I)n_{<{\rm H}>}}{a_{I,{\rm upper}}-a_{{\rm lowest}}}
\end{eqnarray}
where $a_{\rm lowest}=a_{0}n_{\rm out}^{1/3}$, 
$\hat{w}(I)$  is derived by 
integrating the equation
\begin{eqnarray}
\frac{d\hat{w}(I)}{dt}=-\frac{J_{{\rm des}}}{n_{<{\rm H}>}}.
\end{eqnarray}
According to \citet{woi05}, 
the discrete moment $\hat{K}^{{\rm discr}}_{j}$ is evaluated by 
\begin{eqnarray}
\hat{K}^{{\rm discr}}_{j}=\frac{1}{(j+1)a_{0}^{j}}\Biggl\{\displaystyle\sum^{I-1}_{i=1}\hat{w}(i)
\displaystyle\sum^{j}_{k=0}a_{i,{\rm min}}^{k}a_{i,{\rm max}}^{j-k}
+\hat{w}(I)\displaystyle\sum^{j}_{k=0}a_{{\rm lowest}}^{k}a_{I,{\rm max}}^{j-k}\Biggr\}
\end{eqnarray}
for ${\it j}$=0 to 3. The 0--th moment $\hat{K}^{{\rm discr}}_{0}$ and the 3rd moment 
$\hat{K}^{{\rm discr}}_{3}$ give the number of grains per H--element  
and the number of condensible molecules incorporated into grains 
per H--element, respectively. The volume equivalent radius of grains 
$\langle a\rangle$ at a time is defined by 
$\langle a\rangle=a_{0}(\hat{K}^{{\rm discr}}_{3}/\hat{K}^{{\rm discr}}_{0})^{1/3}$.

\section{THE MODEL}

\subsection{Hydrodynamics}

The basic equations for the time--dependent  spherically symmetric 
hydrodynamical model are written in standard finite 
difference form and solved by explicit integration along 
the lines of difference scheme given by \citet{ric67}.  
This modeling method has been adopted by Berlin group
(e.g., \citealt{fle92,win00b,hel00,dre11}).
Actually, this method is time--consuming since the time steps are 
restricted by the Courant--Friedrichs--Levy (CFL) 
stability condition (see \citealt{hof96}). 
However, since the advection term drops out from the equation of motion
in Lagrangian coordinate, we can easily trace the size distribution of 
grains which is one of main subjects of this study.

The continuity equation representing the conservation of mass is
replaced with the definition of gas velocity $v$ given by
\begin{eqnarray}
\frac{\partial r}{\partial t}=v
\end{eqnarray}
where $r$ is the radial position of the Lagrangian mass element at a time $t$.
Under the assumption of position coupling 
between gas and grains,  
the equation of motion is described as 
\begin{eqnarray}
\frac{\partial v}{\partial t}=
-\frac{1}{\rho}\frac{\partial p}{\partial r}-\frac{GM(r)}{r^{2}}(1-\alpha)
\end{eqnarray}
where $p$ is the thermal gas pressure, $\rho$ is the gas density, 
$G$ is the gravitational constant, and $M(r)$ is the mass inside a radius $r$.
In the above equation, $\alpha $ is the ratio of radiation pressure
force to gravity  and expressed as
\begin{eqnarray}
\alpha=\frac{\kappa_{\rm pr}L_{\ast}}{4\pi cGM(r)}
\end{eqnarray}
where $\kappa_{\rm pr}$ is the sum of the flux weighted mean of
mass radiation pressure coefficient for gas $\kappa^{\rm gas}_{\rm pr}$
and that for dust $\kappa^{\rm dust}_{\rm pr}$, $L_{\ast}$  the luminosity, and
$c$ the speed of light.
The mass inside a radius $r$ in the above equations is
replaced with the stellar mass $M_{\ast}$ 
because the mass contained in
the circumstellar envelope is small compared 
to the stellar mass during the calculations 
presented in this paper. 
In the calculations, we adopt the constant gas opacity  of 
$\kappa_{\rm g}$=2$\times 10^{-4}$ cm$^{2}$ g$^{-1}$ \citep{bow88}.
Following the previous works for hydrodynamic models of dust--driven 
wind solved with frequency dependent radiative transfer calculation 
\citep{hof03,woi06,mat10},
we  evaluate the dust opacity $\kappa^{{\rm dust}}_{\lambda}$
at a wavelength $\lambda$ in the Rayleigh approximation.
The optical constants of carbon and SiC grains used in the calculation
are taken from \citet{dra85}\footnote{The Rosseland mean 
and Planck mean absorption coefficients calculated by the 
optical constants of graphite \citep{dra85} are almost the same as those 
calculated by the optical constants of amorphous carbon \citep{rou91} 
used in the recent hydrodynamical calculations (e.g., \citealt{hof03,mat10}). 
Thus, the  difference in the optical constants does not so much affect 
the results of calculations presented in this paper.} 
and \citet{cho85}, respectively. 
The radiation field in the circumstellar envelope 
is derived from the frequency--dependent 
radiative transfer calculation using the variable Eddington
factor method under the assumption of radiative equilibrium.
Then, the gas temperature $T_{\rm gas}$ is given by  
\begin{eqnarray}
\int_{0}^{\infty} \kappa^{\rm gas}_{\lambda,{\rm abs}} J_{\lambda}d\lambda
=\int_{0}^{\infty} \kappa^{\rm gas}_{\lambda,{\rm abs}}
B_{\lambda}(T_{\rm gas})d\lambda
\end{eqnarray}
where the subscript ``abs'' refers to absorption, $J_{\lambda}$
is the mean intensity and $B_{\lambda}(T_{\rm gas})$  the Planck function.
The temperature of $i$--th grain species $T_{{\rm dust},i}$ is 
given by  
\begin{eqnarray}
 \int_{0}^{\infty} \kappa^{{\rm dust},i}_{\lambda,{\rm abs}} J_{\lambda}d\lambda
=\int_{0}^{\infty} \kappa^{{\rm dust},i}_{\lambda,{\rm abs}} B_{\lambda}(T_{{\rm dust},i})d\lambda.
\end{eqnarray}

\subsection{Vibration temperature of SiC cluster}

According to the hydrodynamical models developed so far 
for dust--driven wind from C--rich AGB stars \citep{fle92,hof95},  
the vibration temperature of carbon cluster is set to be equal
to the gas temperature as usual. 
On the other hand, in order to investigate the influence of 
the inverse greenhouse effect suggested by \citet{mcc82} 
for the formation of SiC grains, we consider the two 
cases for the vibration temperature of SiC cluster: 
One is the case that the vibration temperature of SiC cluster is the 
same as the gas temperature and that is referred to as the LTE case hereafter. 
Another is the non--LTE case where 
the vibration temperature is assumed to be the same as the 
temperature of small SiC grains \citep{koz96}.
Assuming that the gas species mainly consisting of H, H$_{2}$, and He   
accommodate with grains 
completely upon colliding, 
and placing a small SiC grain hypothetically at a position in the gas 
flow, we derive the vibration temperature of SiC cluster $T_{\rm v}$  
by  balancing 
the heating with the cooling through the interaction with radiation and
gas as follows;
\begin{eqnarray}
\int\pi Q^{{\rm SiC}}_{\lambda,{\rm abs}}(a_{\rm cl})
[J_{\lambda}-B_{\lambda}(T_{\rm v})]d\lambda 
=\displaystyle\sum_{j}n_{{\rm gas},j}
\left(\frac{k_{\rm B}T_{\rm v}}{2\pi m_{j}}\right)^{\frac{1}{2}}
(C_{{\rm V},j}+\frac{1}{2}k_{\rm B})
\left(T_{\rm v}-T_{\rm gas}\sqrt{\frac{T_{\rm gas}}{T_{\rm v}}}\right)
\end{eqnarray}
where $Q^{{\rm SiC}}_{\lambda,{\rm abs}}(a_{\rm cl})$ is the 
absorption efficiency factor of SiC grain with radius $a_{\rm cl}$ whose value 
is set to be $10^{-3}$ $\micron$ in the calculation,  
and $n_{{\rm gas},j}$, $m_{j}$, and  $C_{{\rm V},j}$ are the number density, 
the mass, and  the specific heat at constant volume
of the $j$--th gas species, respectively. 

\subsection{Modeling procedure}

The models of pulsation--enhanced dust--driven wind
are specified by the six parameters; the stellar mass $M_{\ast}$, 
the stellar luminosity $L_{\ast}$, 
the effective temperature $T_{{\rm eff}}$,  
the C/O ratio, the period of the pulsation $P$, and 
the velocity amplitude of the pulsation $\Delta u_{\rm p}$.
As the first step of the hydrodynamical simulation, 
a dust--free static model atmosphere is 
constructed as follows; 
The initial stellar radius $R_{\ast}$ is given by the
Stefan--Boltzmann law $L_{\ast}=4\pi R_{\ast}^{2}\sigma T_{{\rm eff}}^{4}$ 
where $\sigma$ is the Stefan--Boltzmann constant.
Given a guess value for gas density at $R_{\ast}$,  
the radial profiles of density and temperature are calculated by solving
the equation 
\begin{eqnarray}
\frac{1}{\rho}\frac{dp}{dr}= -\frac{GM_*}{r^2}
\end{eqnarray}
coupled with the radiative transfer calculation with sufficiently fine grids.
The procedure is iterated until the geometrically diluted optical depth
at $R_{\ast}$ converges to 2/3.

The dynamical model calculations including the formation of grains 
are carried out by placing the piston 
at $R_{\rm in0}$ whose position is
two times pressure scale height
around $R_{\ast}$ below from the photosphere of static model 
atmosphere and is defined as the
inner boundary of dynamical model at $t$=0.
Then the radius $R^{\rm dyna}_{\rm in}$ of inner boundary is varied sinusoidally with time as
\begin{eqnarray}
R^{\rm dyna}_{\rm in}=R_{\rm in0}+\Delta u_{\rm p}\frac{P}{2\pi}\sin\left(\frac{2\pi}{P}t\right)
\end{eqnarray}
and the velocity $u^{\rm dyna}_{\rm in}$ at the inner boundary is given by
\begin{eqnarray}
u^{\rm dyna}_{\rm in}=\Delta u_{\rm p}\cos\left(\frac{2\pi}{P}t\right).
\end{eqnarray} 
In the simulation, no mass flow is assumed through the inner boundary.
The radiative flux is assumed to be constant over time and be  
equal to $\pi B(T_{{\rm eff}})=\pi\int_0^{\infty}B_{\lambda}(T_{\rm eff})d\lambda$
at the photosphere of dynamical model $R_{{\rm pht}}$
where the diluted optical 
depth of gas is 2/3, and thus the stellar luminosity 
$L_{\ast}=4\pi R_{{\rm pht}}^{2}\sigma T_{{\rm eff}}^{4}$ 
varies with time.
The piston velocity amplitude $\Delta u_{\rm p}$ 
is increased slowly from 1 cm s$^{-1}$ to the specified value to prevent
the first outwardly moving subsonic wave from growing into a enormous
transient shock, 
following \citet{bow88}.

After the onset of dust formation, the dust--driven wind takes
places.
As the gas expands,  
the spacing between radial grids in the acceleration region becomes wider
and the rezoning procedure necessary for resolving the density
structures of the circumstellar envelope is introduced by adopting 
the method proposed in Fleischer et al.(1992). 
After the outermost zone passes through the outer boundary of dynamical 
model which is placed at 25 $R_{\ast}$ in the simulation, 
the outermost Lagrangian grid is eliminated.
Then, at the outer boundary, we trace the time variations 
of mass--loss rate, gas outflowing velocity, 
dust--to--gas mass ratio,  
and condensation efficiency $f_{\rm C}$ ($f_{\rm Si}$) which is 
defined as the fraction of carbon (silicon) atoms in C--bearing 
molecules except for CO (Si--bearing molecules) locked into 
carbon (SiC) grains. Note that, in the case that all Si atoms 
in Si--bearing molecules are locked into SiC grains, the 
maximum value of $f_{\rm C}$ $\sim$ 0.9 in the model 
with C/O=1.4.

\section{RESULTS}

We adopt the following model parameters in the calculation;
$T_{\rm eff}$=2600 K, $L_{\ast}$=10$^{4}$ 
$L_{\odot}$, $M_{\ast}$=1.0 $M_{\odot}$, C/O=1.4, $P$=650 days, 
and $\triangle u_{\rm p}$=2.0 km s$^{-1}$,  which is a 
typical set of the values in the previous studies 
(e.g., \citealt{fle94,win00b}) and is employed as 
the reference values in the models of C--rich AGB stars that reasonably
reproduce the observed dynamical behaviors \citep{win94,win97,now05}.
The number of radial grids used in the hydrodynamical calculations 
depends on the density structure and is in the range between 800 and
1200 throughout the calculations.
The wavelength region covered in the radiative transfer calculation 
ranges from 0.35 to 120 $\rm{\mu}$m
and the number of wavelength grids is 100.
The simulations have been performed for 180 cycles (stellar pulsation periods). 
In this paper, we neglect the radiation pressure force acting on SiC
grains because of the transparency in the optical 
to infrared region (up to $\sim$ 9 $\micron$) 
\citep{cho85,pit08,hof09}.
 
The formation of SiC grains in the pulsation--enhanced dust--driven wind
is closely related with the formation of carbon dust shell stemming from
the rapid formation of carbon grains in the high
density gas induced by the stellar pulsation 
and the resulting acceleration of dense region caused by the radiation
pressure force acting on the carbon grains.
The carbon dust shell is characterized by the inversions of 
density distributions of gas and carbon grains and plays a 
crucial role in the time evolution of gas temperature and density 
through the backwarming caused by the thermal emission from carbon
grains.  
First, we outline the formation of carbon dust shells in this model,  
and then we show how SiC grains form in the LTE and non--LTE cases.

\subsection{Formation of carbon dust shells}

The carbon dust shell appears in the present model quasi--periodically 
once a stellar pulsation period and is classified into two types by 
the peak value of $f_{\rm C}$ in the shell.
Here we shall overview the formation of each type of the carbon dust
shells  and its dynamics, 
referring Figure 2 which displays 
the radial structure of 
the quantities related to gas dynamics and  
formation of carbon and SiC grains in the LTE case
throughout the period from $t$=166.7 to 168.8 $P$. 

The first type of carbon dust shell 
(hereafter referred to as the 1st type CDS) 
is defined as the high density
region 
with the peak value of $f_{\rm C}\gtrsim$ 0.9, 
and is located 
around 4 $R_{\ast}$ at $t$=167.9 $P$ (see the bottom row in Figure 2). 
The formation of this 1st type CDS actually starts at $t\sim$ 166.7 $P$ 
and the behavior of formation process is almost the same
as that of the 1st type CDS whose formation starts at $t\sim$ 168.8 $P$;
the nucleation of carbon grains takes place in the dust--free region 
ranging from 1.8 to 3.0 $R_{\ast}$ 
whose gas temperature decreases to 1300 K (see  
the radial variation of the nucleation rate of 
carbon grains at $t$=168.8 $P$ and the time variation of the
distribution of the number of carbon grains in this region from  
$t$=168.5 to 168.8 $P$).
Then, the backwarming by thermal emission 
from carbon grains  subsequently formed 
in the outer part of this region not only raises up the gas temperature 
but also evaporates the carbon grains in the inner part 
(see the 3rd rows at $t$=167.0 $P$ and $t$=167.3 $P$). 
However, in the outer part, 
the nucleation and growth of carbon 
grains still proceed and the ratio of radiation pressure force 
to gravity $\alpha$ around 2.5 $R_{\ast}$ exceeds  unity at $t$=167.0 $P$ 
with increasing the volume equivalent radius $\langle a_{\rm C}\rangle$
and the condensation efficiency $f_{\rm C}$ of carbon grains up to 
$\sim$ 0.1 $\rm{\mu}$m and $\sim$ 0.3 (see the 3rd and bottom rows), 
as a result this part is accelerated outwards.
On the other hand, the gas density in the region outer than 
the accelerated region is 
insuffcient to produce abundant enough carbon grains 
to accelerate this region efficiently despite the ongoing nucleation. 
Thus, the radial compression due to the velocity difference 
causes the outer part of the accelerated region to 
accelerate more
through the increase of nucleation and growth rates of 
carbon grains, and results in the 
shell structure; note that the shell structure is not associated with
the pulsation shock but induced by the formation of carbon grains.  
The shell structure becomes prominent 
with decreasing the 
gas density in the region inner than the accelerated region due to the radial 
expansion (see the 2nd rows from $t$=167.3 to 167.6 $P$).  
The formation of 1st type  CDS completes at 
$t$ $\sim$  167.9 $P$. 

The formation of the second type of carbon dust shell 
(hereafter referred to as the 2nd type CDS) defined as 
the high density region with the peak value of 
0.3 $\lesssim f_{\rm C}\lesssim$ 0.7
onsets  around 2.2 $R_{\ast}$ at $t$=168.2 $P$, being triggered 
by the merging of the two consecutive shocks induced by 
the stellar pulsations, which can be seen from the 
time variations of gas velocity and $\alpha$ 
in this region from $t$=167.9 to 168.2 $P$; the carbon 
grains in front of the precedent shock, which 
nucleated at $t$ $\sim$ 166.7 $P$ and 
survived from the evaporation process associated 
with the evolution of 1st type 
CDS starting at $t$ $\sim$ 167.0 $P$, grow up 
from $\langle a_{\rm C}\rangle$=0.05 to 0.1 $\micron$ in the 
region compressed further by the merging with the 
following shock (see the bottom rows from $t$=167.6 to 168.2 $P$). 
After then, 
the nucleation and growth of carbon grains in the compressed 
outer part of this region develops the shell structure with 
increasing $\alpha$,   
while the rapid decrease of gas density caused by the outward 
velocity already exceeding 10 km s$^{-1}$ suppresses 
the growth rate 
of carbon grains and the increase of $f_{\rm C}$. 
The evolution of this carbon dust shell ends up as the 2nd type CDS 
with the peak value of $f_{\rm C}$ $\sim$ 0.7 at $t$ $\sim$ 169.0 $P$. 
The density profile as well as the peak value of $f_{\rm C}$ 
of 2nd type CDS varies from cycle to cycle (e.g., see the 2nd 
type CDS located around 4.5 $R_{\ast}$ at $t$=167.0 $P$). 
The alternating formation of the 1st and 2nd type CDSs is 
sometimes disturbed, in which case the 2nd type CDS 
fails to grow up without increasing the peak value 
of $f_{\rm C}$ up to 0.3. The appearance and periodicity  
of carbon dust shell depend on the stellar parameters 
in a complicated manner as investigated by \citet{fle95}, \citet{hof95}, 
and \citet{dre09,dre11}.
For example, in our model with C/O=1.8, 
only the 1st type CDS forms every stellar pulsation 
period. 

\subsection{Formation of SiC grains in LTE case}

The values of mass--loss rate, 
gas terminal velocity, and  dust--to--gas mass ratio 
averaged over the last sixty pulsation periods at the outer boundary 
are 1.24$\times$10$^{-5}$ $M_{\odot}$ yr$^{-1}$,  
20.4 km s$^{-1}$, and 1.64 $\times$10$^{-3}$, respectively, 
in the hydrodynamical model with the LTE case for the formation of 
SiC grains. As can be seen from Figure 2, 
the nucleation of SiC grains onsets when 
the gas cools down to 1100 K after 
the formation of carbon grains,
and is activated in the temperature range of 800--1000 K. 
The corresponding formation region is located around 3.5
$\sim$ 5.5 $R_{\ast}$ unless the gas temperature is 
raised up higher than 1100 K by the backwarming from the 
precedent 1st or 2nd type CDS. In the formation region, 
 Si$_{2}$C$_{2}$ molecule is more abundant  
than SiC molecule and the nucleation proceeds 
through the reaction (2) rather than the reaction (1). 
In particular,
the formation of SiC grains is 
most efficient in the region around the outer boundary 
of the 1st type CDS 
because of the lower temperature and 
higher gas density, and the nucleation rate per H--element 
reaches the maximum value of $\sim$ 10$^{-25}$ s$^{-1}$ 
at $t$=167.9 $P$. However the subsequent decrease of 
the gas density caused by the accelerated radial expansion depresses 
the nucleation and growth processes of SiC grains; 
the number of SiC grains per H--element is limited to 
smaller than 10$^{-17}$ (see the bottom row at $t$=168.8 $P$), 
and the volume equivalent radius of SiC 
grains $\langle a_{\rm SiC}\rangle$ is typically  
3--6$\times$10$^{-2}$ $\micron$.
In the other regions, the number and size of SiC grains 
are much smaller. 
Thus the depletion of Si--bearing molecules 
does not take place substantially 
in LTE case (see the radial variation of $f_{\rm Si}$ 
in the bottom row throughout the time), and almost all
of SiC grains form around the outer boundary 
of carbon dust shell. 

Figure 3 represents the the radial distribution of gas 
and grains in the region of $r<$ 25 $R_{\ast}$ at $t$=166.8 $P$. 
The radial distribution of $f_{\rm C}$ (the 2nd row) clearly shows that 
the different type of carbon dust shell forms by turns as addressed 
in Section 5.1. 
The carbon dust shells broaden with moving outwards from 
the formation sites because the outer part with more abundant 
carbon grains is accelerated more efficiently than the inner 
part, and the broadening in the outer circumstellar envelope 
of $r>$ 10 $R_{\ast}$ is caused by the pressure gradient.
In addition, the slower moving 2nd type CDS  
seems to be caught up by the faster moving 1st type  
CDS around 21 $R_{\ast}$. 
Note that the shell (shock) structure seen in the radial distribution of
gas density precedes the 
corresponding carbon dust shell in 
the outer circumstellar envelope (see the top row). 
The distributions of $\rho_{\rm SiC}$ (the top row) and 
$f_{\rm Si}$ (the 2nd row) demonstrate that 
SiC grains reside in thin shell around the outer boundary of 
carbon dust shell, reflecting the formation site.
The value of $f_{\rm Si}$ ($\rho_{\rm SiC}$/$\rho_{\rm gas}$) 
and the mass ratio of SiC to 
carbon grains SiC/C are at most 10$^{-6}$ (10$^{-9}$) and 10$^{-6}$, respectively 
(see the lower three rows). 
The value of $f_{\rm C}$ ($\rho_{\rm C}$/$\rho_{\rm gas}$) 
averaged over the last 
sixty pulsation periods at the outer boundary is 0.707 (1.64 $\times$ 10$^{-3}$), 
while the averaged value of $f_{\rm Si}$ 
($\rho_{\rm SiC}$/$\rho_{\rm gas}$) is in the 
order of magnitude of 10$^{-8}$ (10$^{-11}$).
The resulting averaged value of SiC/C is 4.93 $\times$ 10$^{-8}$ 
which is much smaller than the value inferred from the 
radiative transfer models (0.01--0.3). 
 
Figure 4 shows the size distribution  of SiC grains by mass 
averaged over the last sixty pulsation periods at 
the outer boundary. The distribution 
is characterized by the 
single--peaked profile with the
peak around $a$=6.0$\times$10$^{-2}$ $\rm{\mu}$m,  and 
the radius of SiC grains is  limited to be less than 0.1 $\rm{\mu}$m.  
The modal size 
is almost a factor of four smaller than 
the size of presolar SiC grains extracted from the 
Murchison meteorite \citep{ama94}. 
The LTE case with lower C/O ratio resulting 
in lower gas outflow velocity makes the formation of 
SiC grains with radius of 0.2--0.3 $\rm{\mu}$m possible 
as long as the gas outflow exhibits a layered structure with the CDSs. 
On the other hand, regardless of C/O ratio, the amount of SiC grains 
formed in the LTE case is too small to reproduce amount of SiC grains 
inferred from astronomical 
observations. 

\subsection{Formation of SiC grains in non--LTE case}

Figure 5 displays the radial structure of 
the quantities related to gas dynamics and  
formation of grains in the non--LTE case
throughout the period from $t$=166.7 to 168.8 $P$. 
The vibration temperature of SiC 
cluster defined in Section 4.2 and depicted in 
the top row (solid red) is evaluated by placing the small SiC grain with 
$a_{\rm cl}$=10$^{-3}$ $\rm{\mu}$m  hypothetically in the dynamical 
calculations. The difference between the gas temperature $T_{\rm gas}$  
and the vibration temperature $T_{\rm v}$ is negligible in the 
high density region of $\rho$ $>$ 10$^{-12}$ g cm$^{-3}$ 
where the collision with gas dominates the energy balance 
and controls $T_{\rm v}$. As the gas density decreases 
and the collision with gas gets less effective than the 
interaction with the radiation in the energy balance, 
$T_{\rm v}$ deviates from and turns to be lower than $T_{\rm gas}$  
because small SiC grain is almost transparent from the visible to 
infrared region (up to $\sim$ 9 $\micron$) \citep{cho85,pit08,hof09}. 
Also, it should be noted that $T_{\rm v}$
is not affected so much by the backwarming due to the thermal emission 
from the outer carbon dust shell compared with the gas temperature.
Thus, the temperature difference $\Delta T=T_{\rm gas}-T_{\rm v}$ 
is magnified in the less dense region 
inner than the opaque carbon dust shell 
(e.g., see around 2.5 $R_{\ast}$ at $t$=167.3 $P$).

As the gas density decreases, the supersaturation ratio of SiC grains 
quickly exceeds unity with increasing $\Delta T$   
as well as decreasing $T_{\rm v}$.  
The nucleation process of SiC grains can be activated in the 
region with $\rho$ $<$ 10$^{-12}$ g cm$^{-3}$ 
as long as 
$T_{\rm v}\lesssim$ 1200 K and $\Delta T\gtrsim$  200 K,  
even in the region with $T_{\rm gas}\gtrsim$ 1400 K
where the nucleation process of carbon grains is almost  depressed. 
On the other hand, if $T_{\rm v}$ is higher than 
1200 K, the nucleation process of SiC grains is not activated 
regardless of the gas temperature since the abundances of 
molecules responsible for formation of SiC grains 
are not sufficient to make the supersaturation ratio large 
enough for the onset of nucleation.

The active nucleation and growth of SiC grains takes place in the  
infalling gas around 1.8--2.2 $R_{\ast}$ at $t$=166.7 $P$ when 
the gas density drops down to a few times 10$^{-13}$ g cm$^{-3}$, 
$T_{\rm v}$ decreases down below 1200 K and $\Delta T$ increases by 
larger than $\sim$ 200 K , almost concurrently with the onset of
formation of the 1st type CDS as explained in Section 5.1. 
Then, the backwarming from the evolving 1st type CDS raises up 
$T_{\rm v}$ and $T_{\rm gas}$ high enough so as to evaporate 
SiC grains in the inner dense region with $\rho$ $\gtrsim$ 10$^{-13}$ g cm$^{-3}$
at $t$=167.0 $P$. However, in the outer 
less dense region, 
SiC grains grow up to 
$\sim$ 0.3 $\micron$ thanks to the increase of $\Delta T$ and 
the condensation efficiency $f_{\rm Si}$ reaches 0.1.  
In the region around 2.5 $R_{\ast}$, the nucleation of SiC grains 
proceeds, and at $t$=167.3 $P$ the increase of  $f_{\rm Si}$ is 
distinguishable while carbon grains evaporate due to the 
backwarming, which continues up to $t\sim$167.5 $P$ 
(see the 3rd and bottom rows). 
On the other hand, 
in the region around 2 $R_{\ast}$, 
the evaporation of SiC grains turns to the growth at $t$=167.3 $P$ 
with weakening the backwarming as the 1st CDS moves outwards, 
and then the increase of gas density up to 10$^{-13}$ g cm$^{-3}$ 
caused by the pulsation shock 
leads to the efficient nucleation and growth of 
SiC grains, and  $f_{\rm Si}$ increases up to $\sim$ 0.7 quickly 
until  $t$=167.6 $P$.  
Then, the efficient growth of SiC grains in the region 
compressed by the merging of consecutive shocks starting from 
$t$ $\sim$ 167.9 $P$ 
results in the complete consumption of Si--bearing molecules 
($f_{\rm Si}$=1.0) around 
$t\sim$168.1 $P$, 
and $\langle a_{\rm SiC}\rangle$ in this 
region is typically larger than $\sim$ 0.2 $\micron$,  
in contrast to the LTE--case. The region with the 
condensation efficiency $f_{\rm Si}$ $\sim$ 1.0 is referred to as 
the Si--depletion region hereafter. 
Concurrently, 
the 2nd type CDS formed and developed in this region 
stretches the Si--depletion region outwardly (see the radial 
distribution of $f_{\rm Si}$ in the bottom row from 
$t$=168.2 to 168.8 $P$),   
while the backwarming from the 2nd type 
CDS makes $T_{\rm v}$ higher than 1200 K  and 
depresses the nucleation of SiC grains 
in the region inner than 
the Si--depletion region for a while
(see $T_{\rm v}$ in the top row at $t$=168.5 $P$).
It should be noted that the 
2nd type CDS moves outward more slowly than the 1st type 
CDS as mentioned in Section 5.2.
Thus, the Si--depletion region with slower outward 
velocity is finally caught up and pushed 
by the 1st type CDS as 
can be seen from the time evolution of the 
Si--depletion region spreading over 
from 3.5 to  5.5 $R_{\ast}$ 
at $t$=167.3 $P$ which formed two cycles before.

As mentioned above, the formation of Si--depletion region 
onsets in the infalling gas in front of the coming pulsation shock,  
and the region moves outward with the formation of 2nd type CDS 
triggered by the merger with the next pulsation  shock. 
Thus, the Si--depletion region forms once every two pulsation 
periods. 
Also, even in the other regions,  
the resulting value of $f_{\rm Si}$  often exceeds 0.1. 
Thus, in contrast to the very small value of at most 10$^{-6}$ 
in LTE case, a large amount of SiC grains can 
condense through the inverse greenhouse effect in the non--LTE 
case.

Figure 6 presents the radial variations of gas and grains 
at $t$=176.3 $P$. It seems that SiC grains are concentrated 
around the outer boundaries of 1st type CDSs  
in the density distribution 
(see the top row) because the shock associated with the 
formation of carbon dust shell sweeps up the gas. 
Reflecting the variation of gas density, 
the radial distribution of $\rho_{\rm SiC}$ does not 
always corresponds to the Si--depletion region 
shown in the 2nd row. It 
should be noted that the extended 
Si--depletion region does not develop without 
the 2nd type CDS,  
which can be demonstrated from an example 
of the failure of alternating formation of both types of 
carbon dust shells located at 18 $R_{\ast}$ where  Si--depletion region 
restricted in a narrow region resides just ahead of the 
1st type CDS formed about 1 cycle after from the 
formation of the Si--depletion region 
(see the 2nd row). 
Apart from the 1st type CDS located at  18 $R_{\ast}$, 
the extended SiC dust shell corresponding to the 
Si--depletion region between 1st and 2nd type CDSs 
can be clearly recognized in the radial distribution of 
$\rho_{\rm SiC}/\rho_{\rm gas}$ (see the 3rd row where 
$\rho_{\rm SiC}/\rho_{\rm gas}$=8.46$\times$10$^{-4}$ 
corresponding to $f_{\rm Si}$ = 1.0). The radial distribution 
of $f_{\rm Si}$ shows that  the Si--depletion region 
stretched over by the slower moving 2nd type CDS narrows with 
moving outwardly, being compressed by the faster moving 
1st type CDS. Thus, in the outer circumstellar envelope,  
the radial width tends to be of 
the order of the stellar radius 
(see $f_{\rm Si}$ around 24 $R_{\ast}$ 
where the 2nd type CDS almost merges with the 1st type CDS) and  
the narrower Si--depletion region seems to reside 
around the outer boundary of the 1st type CDS. 

The radial distribution of the mass ratio SiC/C has a 
sharp peak 
in the region just ahead of 
the outer boundary of 1st type CDS, and decrease 
by 0.1 (10$^{-4}$) 
toward the outer boundary of 
precedent 2nd (1st) type CDS (see the bottom row). 
The peak value exceeds ten in the inner circumstellar 
envelope, and decreases with increasing $r$ because of 
the subsequent growth of carbon grains. However the 
value of SiC/C in the Si--depletion region still keeps 
larger than 0.2 with the peak value exceeding unity.

The resulting values of mass--loss rate, 
terminal velocity, dust--to--gas mass ratio averaged over 
the last sixty stellar pulsation periods are  
1.10$\times$10$^{-5}$ $M_{\odot}$ yr$^{-1}$,  19.5 km s$^{-1}$, and 
1.85$\times$10$^{-3}$, respectively, for the non--LTE case.  
The reduction of terminal velocity compared to the LTE 
case reflects the somewhat weakened acceleration resulting from
the reduction of amount of carbon grains 
due to the incorporation of C--bearing gas species into SiC grains;  
on average 4.40 \% of carbon atoms 
in the condensible C--bearing molecules is incorporated into the SiC grains.
The averaged value of $\rho_{\rm C}/\rho_{\rm gas}$ ($f_{\rm C}$) is 
1.51$\times$10$^{-3}$ (0.652), while 
the value of $\rho_{\rm SiC}/\rho_{\rm gas}$ ($f_{\rm Si}$) is 
3.41 $\times$10$^{-4}$ (0.403).   
The resulting averaged value of SiC/C is 0.226 
and is close to the upper value derived 
from the radiative transfer models. 
The size distribution of SiC grains by mass 
averaged over the last sixty stellar pulsation periods at 
the outer boundary is depicted in Figure 7. 
The size distribution shows the broad profile with the 
mean radius of 0.3 $\micron$, 
in contrast to the sharp and single--peaked 
size distribution in the 
LTE case presented in Figure 4. 
In the size range from 0.03 to 4.0 $\micron$,  
80 \% of SiC grains by mass populate in the radii 
between 0.2 and 1.0 $\micron$:  
The distribution 
well covers the size range of presolar SiC grains found in the
Murchison meteorite.  
Thus the hydrodynamical model with the non--LTE case 
for the formation of SiC grains that 
the vibration temperature of SiC cluster
is assumed to be the same as the temperature  
of small SiC grains reasonably reproduces the 
amount of SiC grains inferred from astronomical observations and 
the size of presolar SiC grains extracted from the Murchison meteorite.

\section{DISCUSSION}

In the calculations of hydrodynamical models 
presented in Section 5, we assume that 
SiC is the starting molecule for the nucleation process
of SiC grains, excluding the process starting 
from the two--body 
reaction of Si$_{2}$C$_{2}$ molecules. This may underestimate 
the amount of SiC grains in the LTE case since Si$_{2}$C$_{2}$ molecule 
is more abundant in the lower temperature region than SiC molecule 
as shown in Figure 1 (left). In the LTE case where 
carbon grains condense in prior to SiC grains, 
the nucleation process starting from the two--body reaction 
of Si$_{2}$C$_{2}$ molecules dominates the nucleation rate 
in the region around 6--8 $R_{\ast}$  whose gas temperature 
ranges from 600  to 800 K. However the gas density in the region 
of $\rho\sim$10$^{-16}$ g cm$^{-3}$ is too low to increase the amount 
of SiC grains; even if this process is included in the calculation, 
the resulting values of both $f_{\rm Si}$ and SiC/C are limited to
10$^{-4}$ although the nucleation rate is much larger than 
that starting from SiC molecule. In addition, this nucleation process
produces the population of grains whose radii are much smaller than 
6.0 $\times$ 10$^{-2}$ $\micron$, and make the modal radius smaller 
than 3.0 $\times$ 10$^{-2}$ $\micron$. Thus the nucleation process 
starting from two--body reaction of Si$_{2}$C$_{2}$ molecules 
does not change the conclusion that the hydrodynamical models with 
the LTE--case cannot reproduce the amount of SiC grains around 
C--rich AGB stars inferred 
from the astronomical observation and the size range derived from 
the analysis of presolar grains.

In the non--LTE case, this 
nucleation process dominates the formation of SiC grains 
in the region of $r>$4 $R_{\ast}$, and is activated 
particularly in the outer part of 
the carbon dust shell. 
Although the number of SiC grains 
produced through this process reaches about 10 times that 
produced through the nucleation starting from SiC molecule, 
the typical radius of SiC grains formed in the region with 
low gas density is limited to 
3$\times$10$^{-3}$--6$\times$10$^{-2}\micron$.  
Thus, the inclusion of this nucleation process does not affect the 
size distribution of SiC grains whose radius is larger 
than 0.1 $\micron$. Also it is true for the amount of SiC 
grains since the averaged $f_{\rm Si}$ increases by only a few 
percent. Thus we conclude that the exclusion of nucleation 
process starting from the two--body reaction of Si$_{2}$C$_{2}$ molecules 
does not affect the result of calculations in the non--LTE case, 
except for the size distribution in the range 
of $a_{\rm SiC}<$ 6$\times$10$^{-2}\micron$. 

The sticking probability set to be unity for nucleation 
and growth processes of SiC grains may overestimate the 
size and amount of SiC grains formed in the non--LTE case. 
Figure 8 shows the time averaged size distributions by mass 
of SiC grains for given values of $\alpha_{\rm s}$. 
In the range of $\alpha_{\rm s}$=0.1--1.0, the mass 
fraction of large sized grains with $a$ $>$ 1.0 $\micron$ 
decreases significantly and the profile becomes narrow with 
decreasing $\alpha_{\rm s}$, while the 
reduction of $\alpha_{\rm s}$ does not affect the mass fraction 
of small--sized grains with $a$ $\leq$0.1 $\micron$. 
The modal radii are 0.30 and 0.22 $\micron$ 
for  $\alpha_{\rm s}$=0.5 and 0.1, respectively.
Although the SiC grains with radii larger than 1.0 $\micron$ 
do not form for the case of $\alpha_{\rm s}$=0.1, 
the hydrodynamical models in the non--LTE case 
reasonably reproduce the size range
of presolar SiC grains extracted from the Murchison  
meteorite \citep{ama94} as long as $\alpha_{\rm s}\gtrsim$ 0.1. 
The averaged value of SiC/C decreases with decreasing $\alpha_{\rm s}$; 
the values are 0.199 and 
9.77$\times$10$^{-2}$ for $\alpha_{\rm s}$=0.5 
and 0.1, respectively, which is within the range of 
the value inferred from the radiative transfer models.  
Although the reduction of $\alpha_{\rm s}$ changes the 
radial distribution of SiC grains substantially, we conclude that 
the hydrodynamical models in the non--LTE case cover the ranges of size 
derived from the analysis of presolar grains and 
amount evaluated from  the radiative transfer 
models unless the sticking probability $\alpha_{\rm s}\lesssim$0.1.

The mass ratio of SiC/C inferred from the 
radiative transfer models depends on
the models used in the calculation; 
the evaluated values 
for C--rich AGB stars with the emission feature of SiC grains are 
in the range of 0.01--0.29 \citep{lor94}, 0.01--0.15 \citep{gro95,gro98}
, and 0.1--0.3 \citep{bla98}.
The smaller value by \citet{gro95} and \citet{gro98} 
is considered to be ascribed to the  grain model; the temperature 
of composite grain consisting of carbon and SiC used in their model 
is higher than that of isolated SiC grains 
used in the other models as is shown in \citet{bla98}, since carbon 
grains are more absorptive in the  visible to NIR region than  SiC 
grains. Therefore, in order to reproduce the observed SEDs, 
if the inner boundary of dust layer is placed close to the 
photosphere, the high temperature of composite grains 
causes the value of SiC/C to reduce. 
Also, the smaller value of 0.01--0.06 derived by \citet{lor94}
for C--rich AGB stars showing SiC emission feature 
with optically thick shells ($\tau_{\rm 1\mu m} > 3.0$) 
would arise from the optical constant of SiC grains 
used in the models; the 
radiative transfer models usually employ the optical 
constants derived from the SiC powder synthesized in the 
laboratory \citep{bor85,peg88}.  
The absorption coefficients in 
visible to infrared region (up to $\sim$ 9 $\micron$) is much larger than 
the crystalline SiC \citep{cho85,pit08,hof09}, which 
results in the higher temperature of SiC grains and 
the smaller SiC/C. In addition, the 
mid--infrared observation of C--rich AGB stars by 
\textit{Spitzer}
shows that the 30 $\micron$ emission feature attributed to 
MgS grains becomes prominent with decreasing the strength 
of SiC emission feature in the stars with mass--loss rate 
larger than 10$^{-6}$ $M_{\odot}$ yr$^{-1}$ \citep{lei08}.
\citet{zhu08b} demonstrates that 
the absorption efficiency around 11.3 $\micron$
of grains consisting of a SiC core and 
a MgS mantle decreases with increasing the 
volume fraction of MgS mantle. 
Therefore, if it is true, the radiative transfer model 
could underestimate the value of SiC/C 
without including SiC grains coated by MgS mantle. 
Thus, the mass ratio of SiC to carbon grains derived from 
the the radiative transfer models depends on the optical 
constants of SiC as well as the model of grains 
used in the calculations, and the value of SiC/C 
up to $\sim$ 0.23 calculated in the non--LTE 
case for C--rich AGB stars  
is considered to be not always in conflict with the 
values derived from SED fittings.

\section{SUMMARY}

The formation of SiC grains around C--rich AGB stars is 
investigated for the first time in the 
framework of hydrodynamical model for the pulsation--enhanced 
dust--driven wind. 
We formulate 
the nucleation and growth processes of SiC grains, 
considering that SiC grains nucleate and grow homogeneously 
starting from SiC molecule. In the calculations, the two cases 
are considered for the nucleation process of SiC grains; the 
LTE case in which the vibration temperature $T_{\rm v}$ of SiC 
cluster is equal to the gas temperature and the non--LTE case in 
which  $T_{\rm v}$ is assumed to be the same as the temperature 
of small SiC grain whose radius is 10$^{-3}$ $\micron$. 
On the other hand, the vibration temperature of carbon cluster 
is considered to be the same as the gas temperature as is 
usually assumed in the hydrodynamical models for C--rich AGB 
stars.

The results of calculations for the model parameters 
$M_{\ast}$=1.0 $M_{\odot}$, $L_{\ast}$=10$^{4}$ $L_{\odot}$, 
$T_{\rm eff}$=2600 K, C/O ratio=1.4, $P$=650 days, and 
$\Delta u_{\rm p}$=2.0 km s$^{-1}$ are summarized as follows: 
In the LTE case, SiC grains form in the accelerated and low--density 
outflowing gas after the formation of carbon grains. The 
resulting time averaged 
number fraction of Si atoms locked into SiC grains 
(condensation efficiency $f_{\rm Si}$) of at most 10$^{-8}$ 
is too small to reproduce the amount of SiC grains inferred from 
the astronomical observations and the radius is limited to less than 
0.1 $\micron$. On the other hand, in the non--LTE case, the time 
averaged mass ratio of SiC to carbon grains ranges from 0.098 to 
0.23 for the sticking probability $\alpha_{\rm s}$=0.1--1.0, which 
is not in conflict with the value of 0.01--0.3 inferred from the 
radiative transfer models. The time averaged size distribution of 
SiC grains by mass with the peak at radius of 0.2--0.3 $\micron$ well 
covers the size range of presolar 
SiC grains extracted from the Murchison meteorite. Thus we 
conclude that the so called inverse greenhouse effect plays 
a crucial role in the formation process of SiC grains in the 
pulsation--enhanced dust--driven winds from C--rich AGB stars, 
apart from the validity of the underlying assumption made in the 
non--LTE case. The non--LTE effect should be explored for the 
formation process of dust grains in astrophysical environments. 

The hydrodynamical model provides 
the two types of carbon dust 
shells; the 1st (2nd) type CDS with the peak value of 
$f_{\rm C}\gtrsim$0.9 (0.3 $\lesssim f_{\rm C}\lesssim$ 0.7). 
The formation region and the resulting radial distribution 
of SiC grains in the hydrodynamical model with non--LTE case 
is closely related with the formation and dynamics 
of carbon dust shell,  especially through the density 
enhancement and the backwarming effect. In the inner circumstellar envelope, the SiC dust 
shell can be discriminated from the 1st type CDS through 
the extended Si--depletion region caused by the 2nd CDS 
formed in the outer part of active SiC formation region. However, 
in the outer circumstellar envelope of $r>$20 $R_{\ast}$, 
the SiC dust shell is localized around the outer boundary 
of 1st type CDS. 
Not only the formation region and the radial 
distribution but also the amount and the size distribution 
of SiC grains could heavily depend on the C/O ratio 
as well as the other model parameters; 
for instance, in the non--LTE case with 
$M_{\ast}$=2.0 $M_{\odot}$, the mass ratio of SiC to gas 
and the size of SiC are almost the same as those with 
$M_{\ast}$=1.0 $M_{\odot}$ while the mass loss rate  
decreases by 
a factor 5. Anyway more comprehensive study covering a wide range 
of model parameters is necessary for revealing the amount and 
the size distribution of SiC grains formed around C--rich 
AGB stars. Also, the difference between the radial distributions of 
carbon and SiC grains could greatly influence 
the appearance of the spectral feature attributed to SiC grains. 
These aspects will be investigated in the forthcoming papers.

%

\acknowledgments

We are grateful to the anonymous referee for valuable comments that 
improved the manuscript. We thank Dr. K. Ohnaka for his critical 
reading of the first draft and useful comments.
This research has been partly supported by the Grant--in--Aid for 
Scientific Research of the Japan Society for the Promotion of Science 
(18104003, 20340038).

\clearpage



\onecolumn

\begin{figure}
\includegraphics[angle=0,scale=1.1]{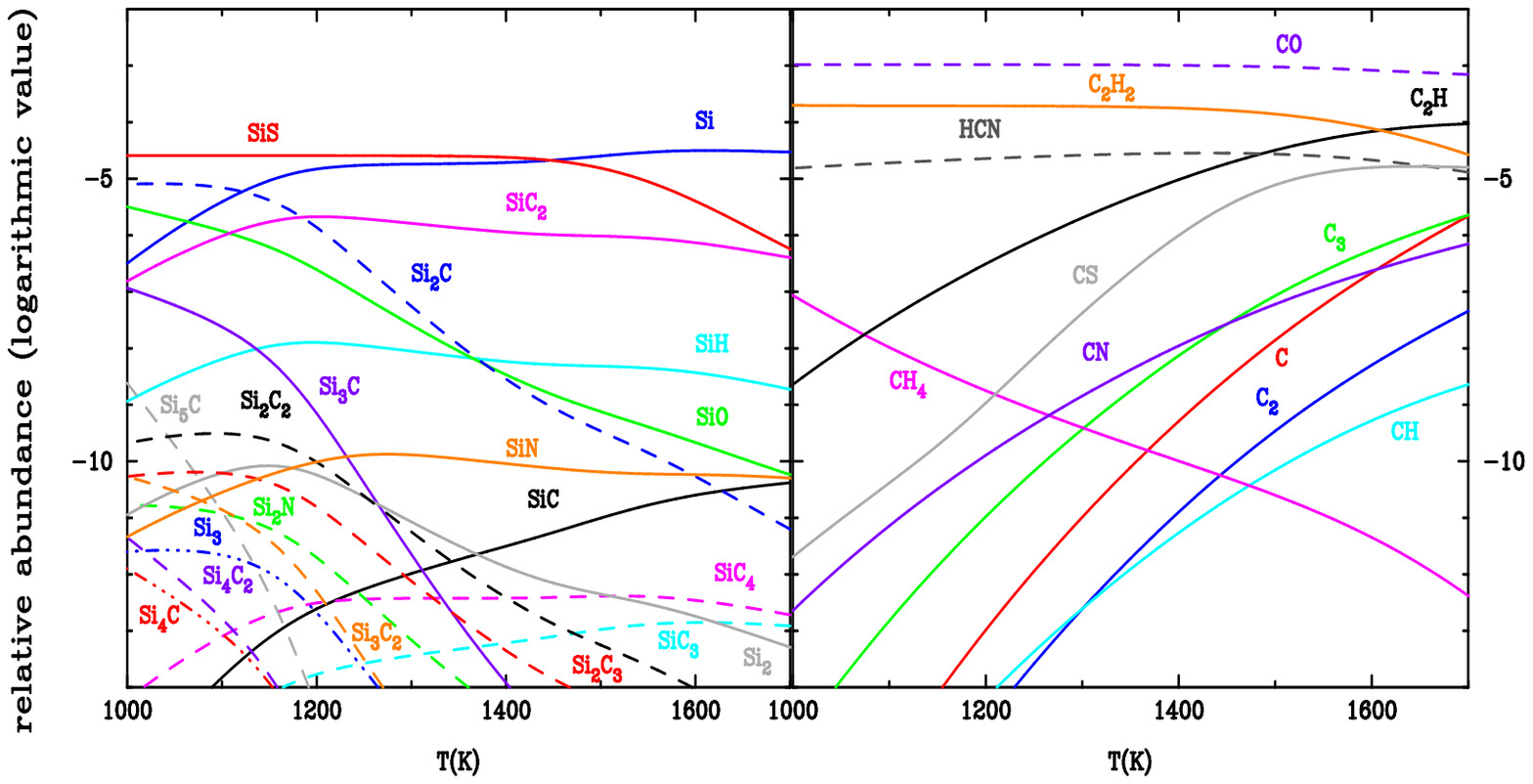}
\vspace{0.1cm}
\caption{Relative abundances of Si--bearing molecules (left)
and C--bearing molecules (right) 
in chemical equilibrium in the gas with the
density $\rho=$ 10$^{-13}$ g cm$^{-3}$ and C/O ratio = 1.4. 
\label{fig1}}
\end{figure}

\clearpage

\newpage

\begin{figure}

\includegraphics[angle=-90,scale=.63]{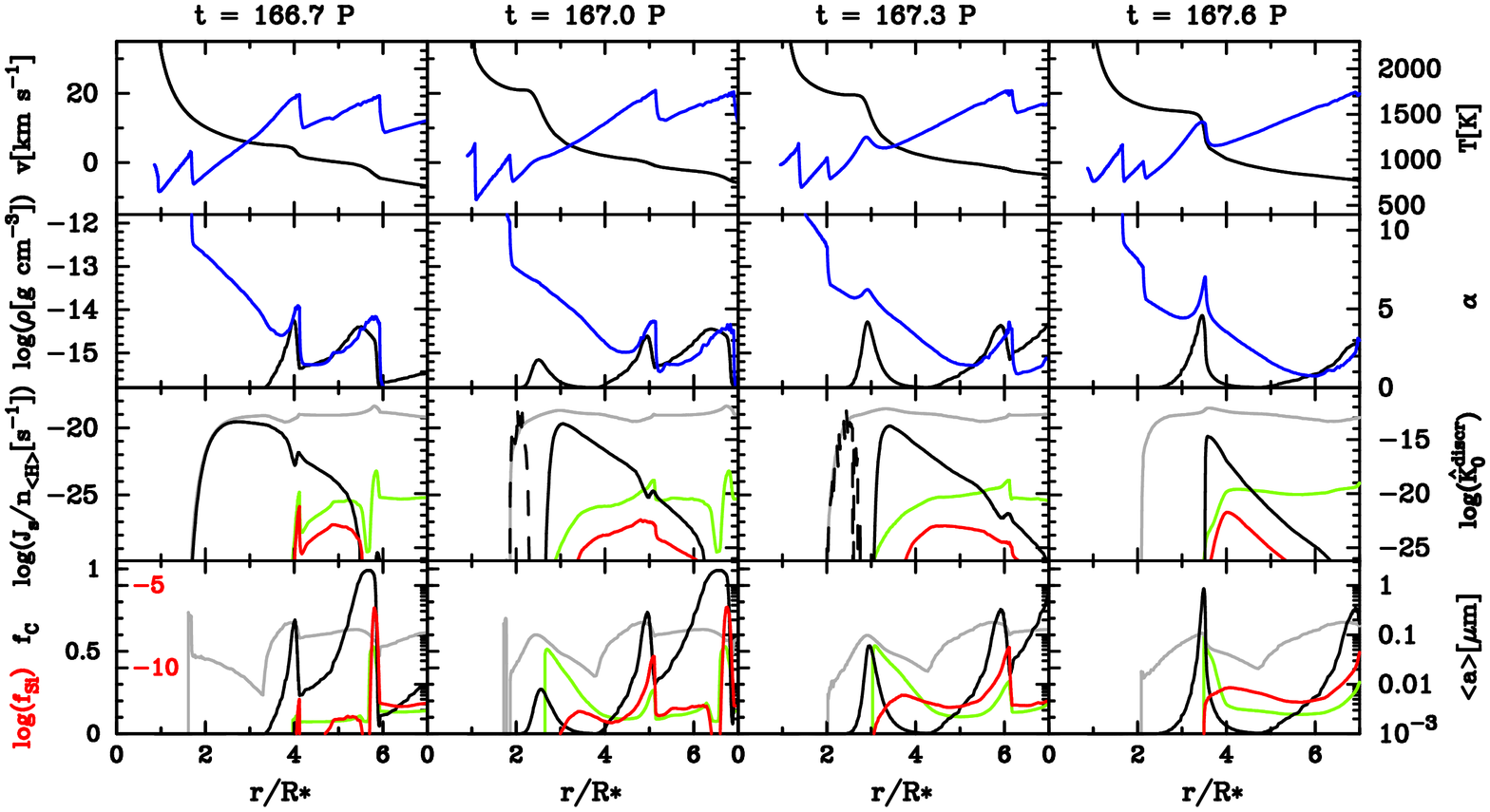}
\vspace{1.0cm}

\includegraphics[angle=-90,scale=.63]{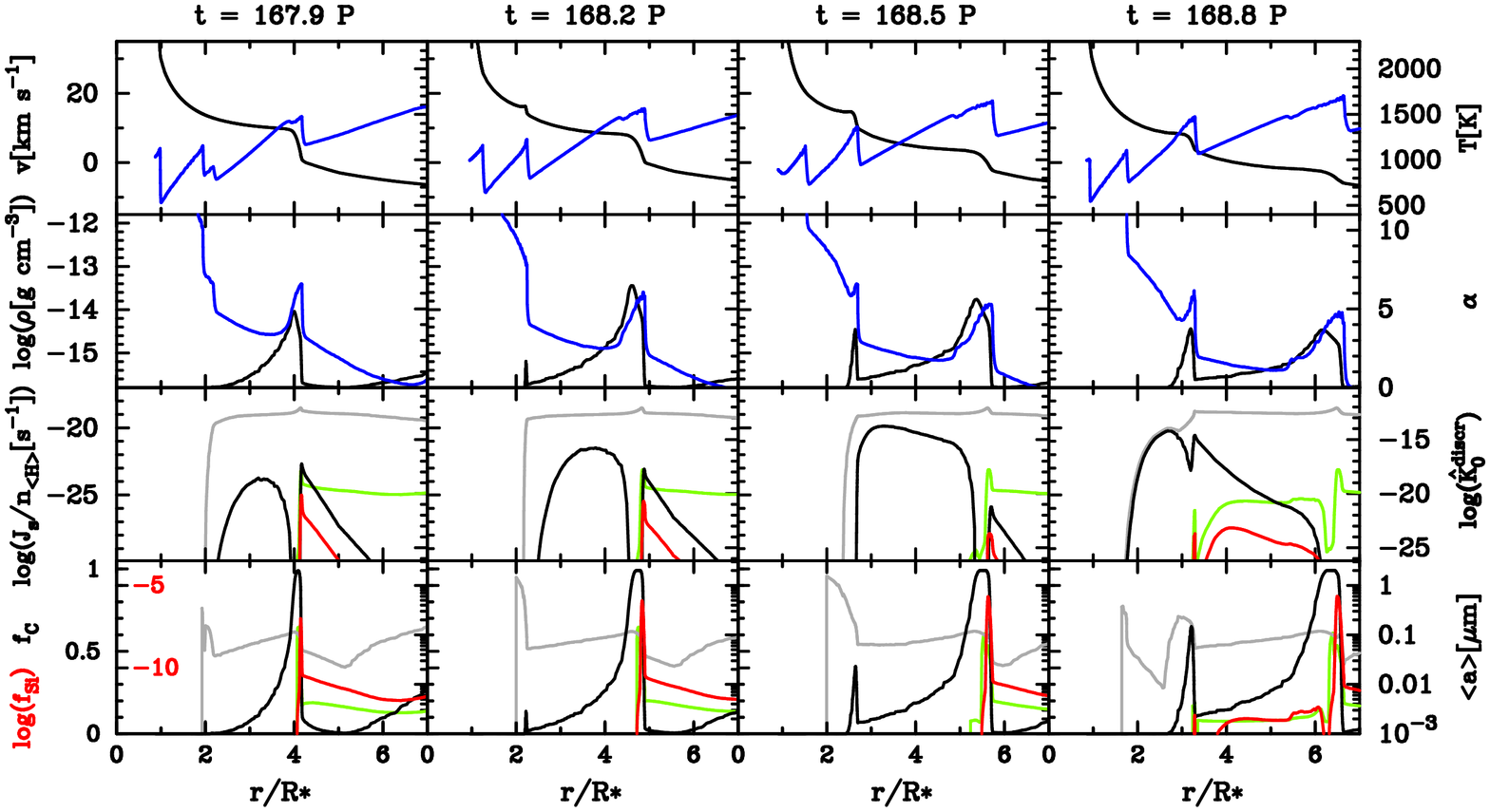}
\vspace{0.5cm}

\caption{Radial structure in LTE case for $t$=166.7 to 167.6 $P$ (top panel)
and for $t$=167.9 to  168.8 $P$ (bottom panel).
Top row; the gas velocity $v$ (blue) and  
the gas temperature $T_{\rm gas}$ (black).   
2nd row; the gas density $\rho$ (blue)
and the ratio of radiation pressure force to gravity $\alpha$ (black). 
3rd row; the nucleation rates per H--element $J_{\rm S}/n_{<\rm H>}$ 
of carbon (black solid) and SiC (red) grains, 
the destruction rate per H--element $J_{\rm des}/n_{<\rm H>}$ of
carbon grains (black dashed), 
and the numbers per H--element $\hat{K}_{0}^{\rm discr}$ of 
carbon (grey) and SiC (green) grains. 
Bottom row; the condensation efficiencies $f_{\rm C}$ (black) 
and $f_{\rm Si}$ (red), 
and the volume equivalent radii $\langle a\rangle$ of carbon 
(grey) and SiC (green) grains.\label{fig2}}
\end{figure}

\clearpage

\newpage

\begin{figure}
\vspace{-2.0cm}
\hspace{0.6cm}
\includegraphics[angle=-90,scale=.85]{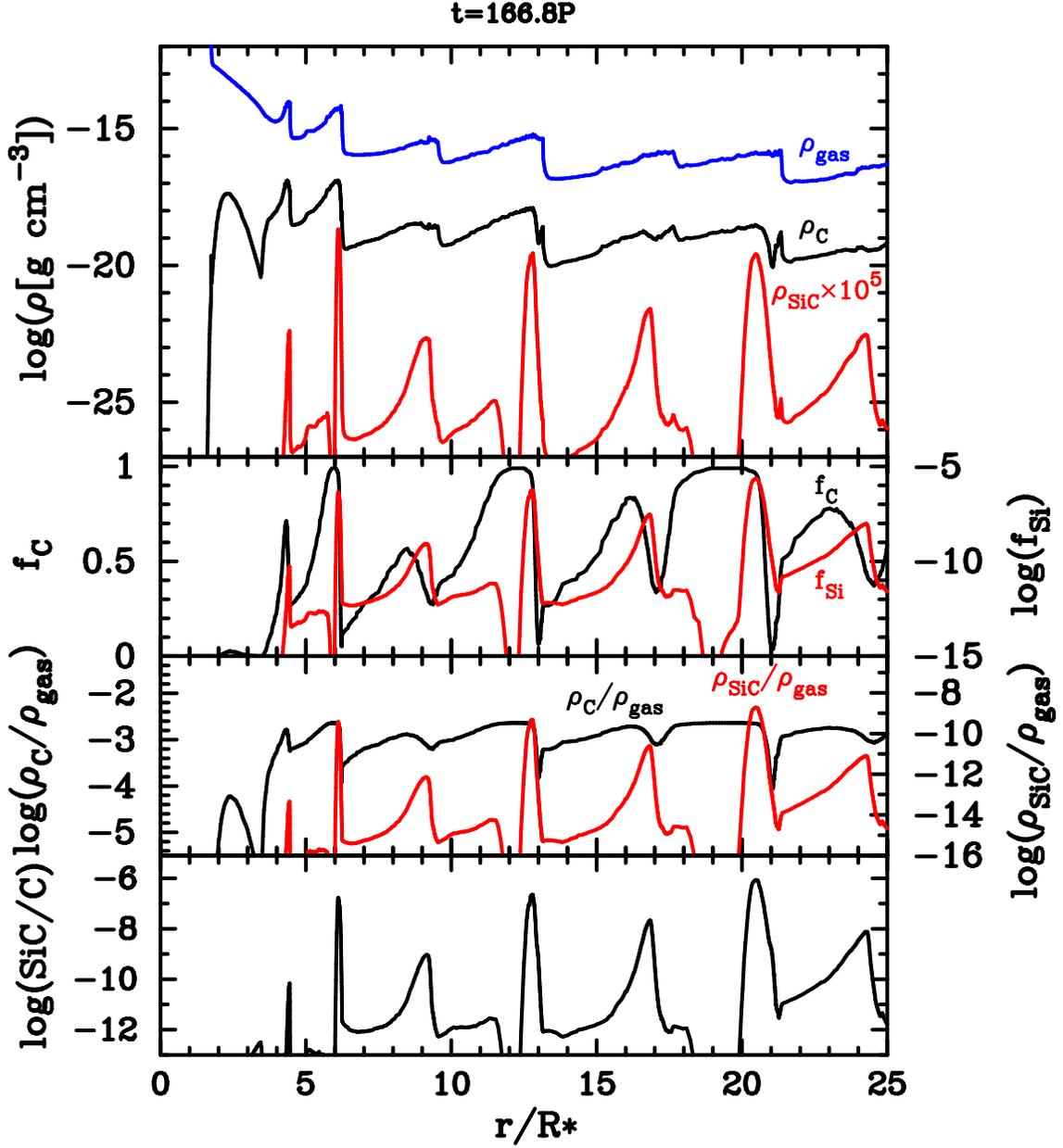}
\vspace{1.2cm}
\caption{Radial distributions of gas and  grains  at $t$=166.8 $P$ in
 the LTE case. Top row; the mass densities of gas 
(blue, $\rho_{\rm gas}$),  carbon (black, $\rho_{\rm C}$) 
and SiC (red, $\rho_{\rm SiC}$) grains. 2nd row; the condensation 
efficiencies $f_{\rm C}$ (black) 
and $f_{\rm Si}$ (red). 3rd row; the density ratios of 
$\rho_{\rm C}/\rho_{\rm gas }$ (black) and  
$\rho_{\rm SiC}/\rho_{\rm gas }$ (red).  
Bottom row; the mass ratio of SiC to carbon grains SiC/C.\label{fig3}}
\end{figure}

\newpage

\begin{figure}
\includegraphics[angle=-90,scale=.55]{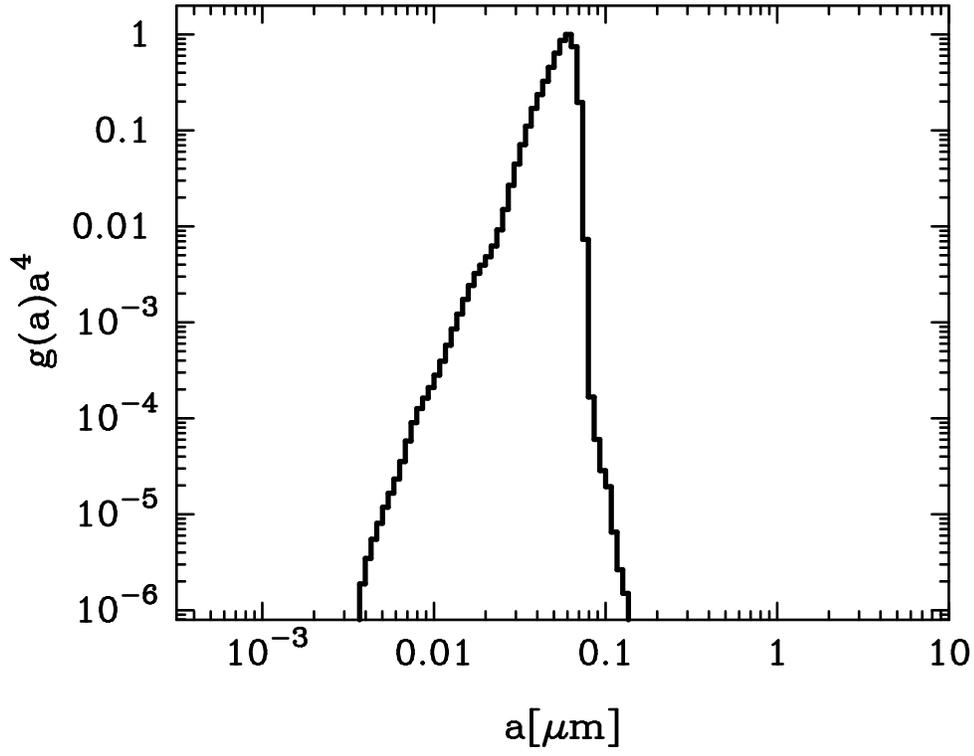}
\vspace{1.5cm}
\caption{Size distributions of SiC grains by mass 
averaged over the last sixty stellar pulsations at the 
outer boundary in the LTE case. Note that $g(a)$ is the 
size distribution function and the value of $g(a)a^{4}$ 
is normalized to the maximum value. 
\label{fig4}}
\end{figure}

\clearpage

\newpage

\begin{figure}

\includegraphics[angle=-90,scale=.63]{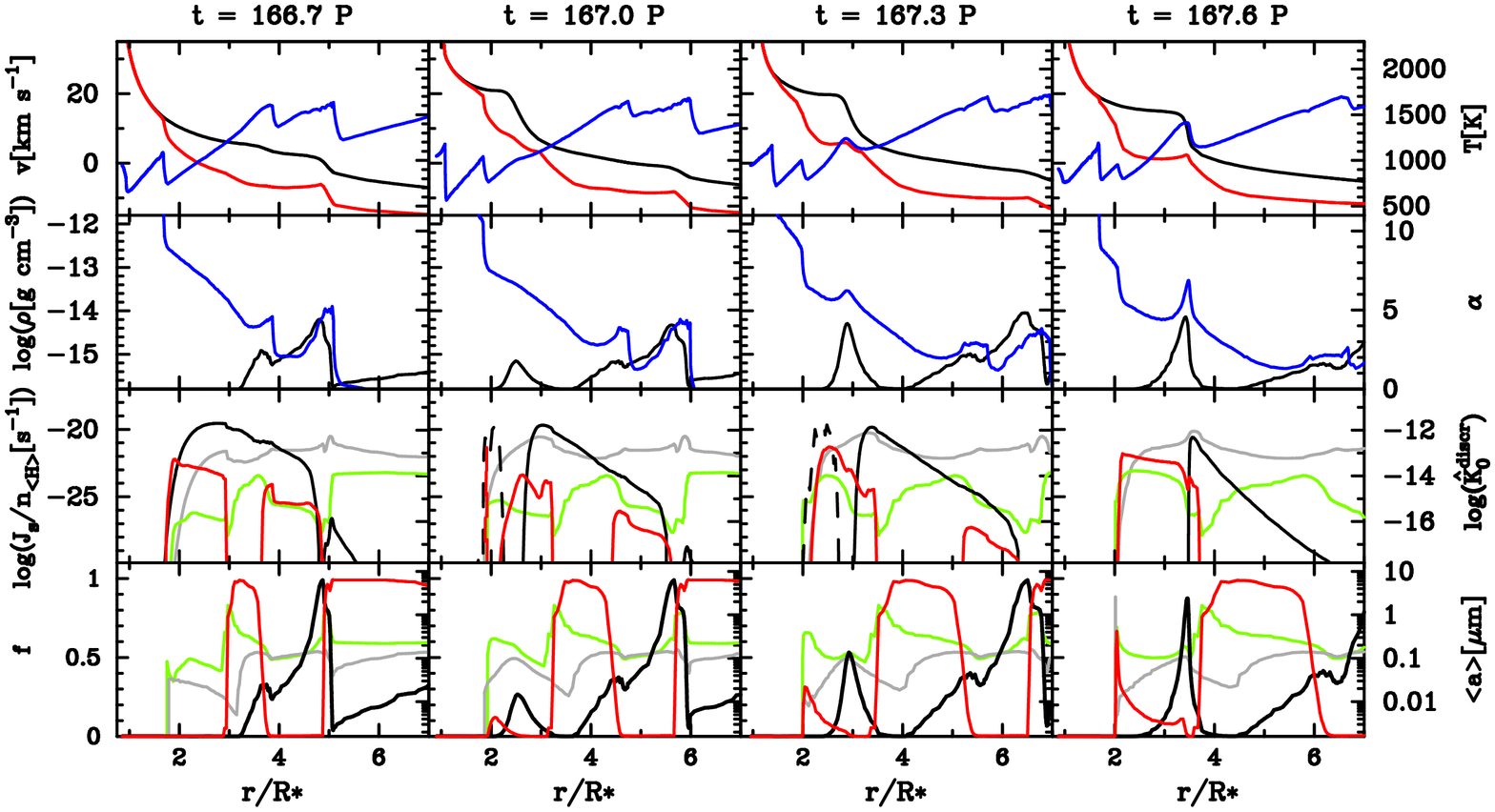}
\vspace{1.0cm}

\includegraphics[angle=-90,scale=.63]{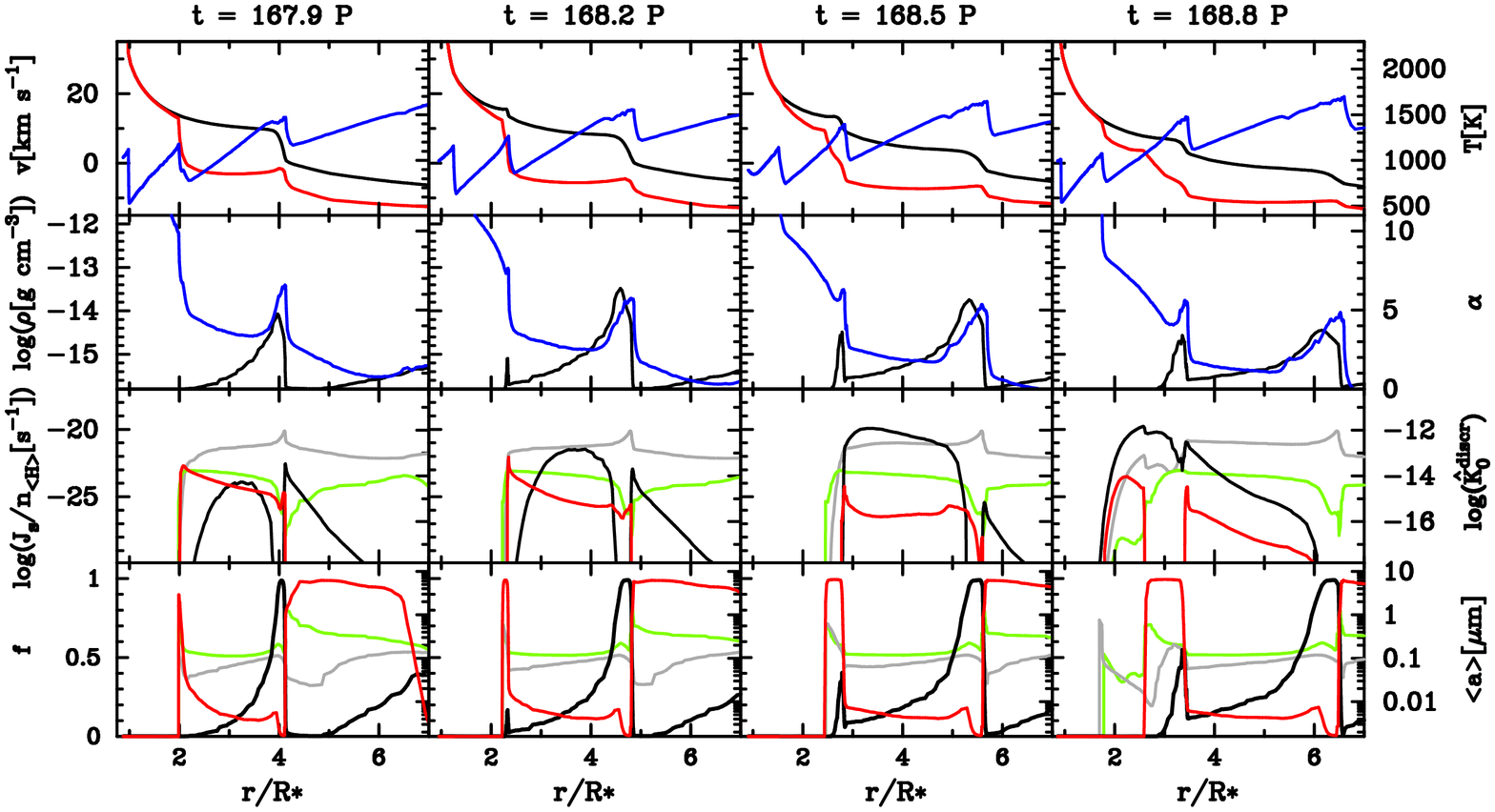}
\vspace{0.5cm}

\caption{Same as Figure 2, but for  the non--LTE case.
Note that the red line in the top rows 
denotes the vibration temperature $T_{\rm v}$   
and the red--dashed line around 1.9 $R_{\ast}$ 
in the 3rd row at $t$=167.0 $P$ 
depicts the destruction rate of SiC grains  per H--element $J_{\rm des}/n_{<\rm H>}$.
 \label{fig5}}
\end{figure}

\clearpage

\newpage

\begin{figure}
\vspace{-2.0cm}
\hspace{2.0cm}
\includegraphics[angle=-90,scale=.75]{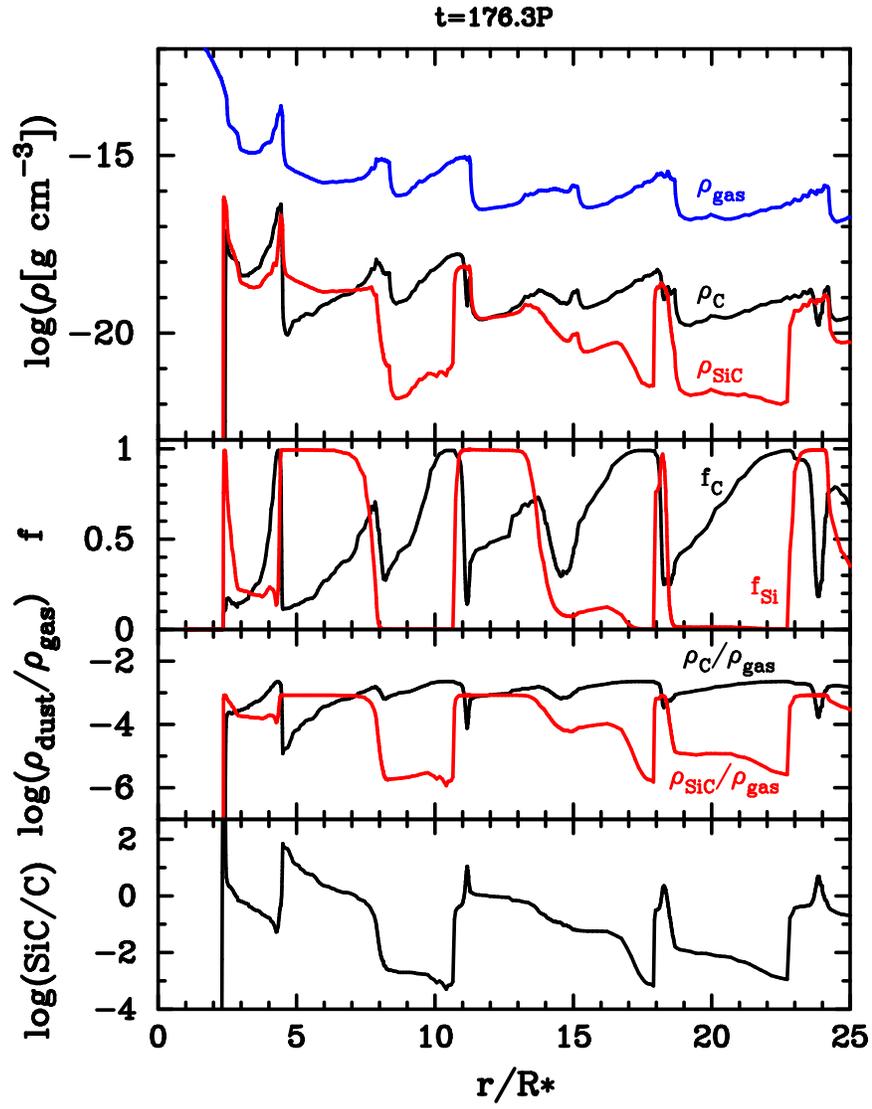}
\vspace{1.2cm}
\caption{Same as Figure 3 but for $t$=176.3$P$ in the non--LTE case.\label{fig6}}
\end{figure}

\newpage

\begin{figure}
\includegraphics[angle=-90,scale=.55]{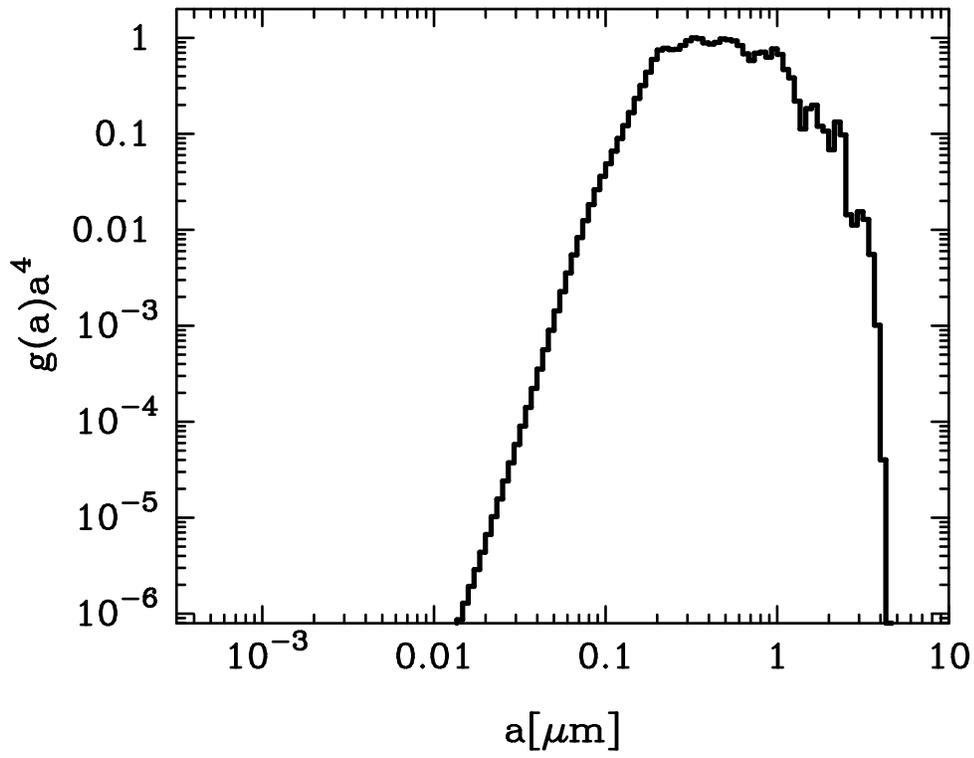}
\vspace{1.5cm}
\caption{Same as Figure 4, but for the non--LTE case. 
\label{fig7}}
\end{figure}

\clearpage

\newpage

\begin{figure}
\includegraphics[angle=0,scale=1.20]{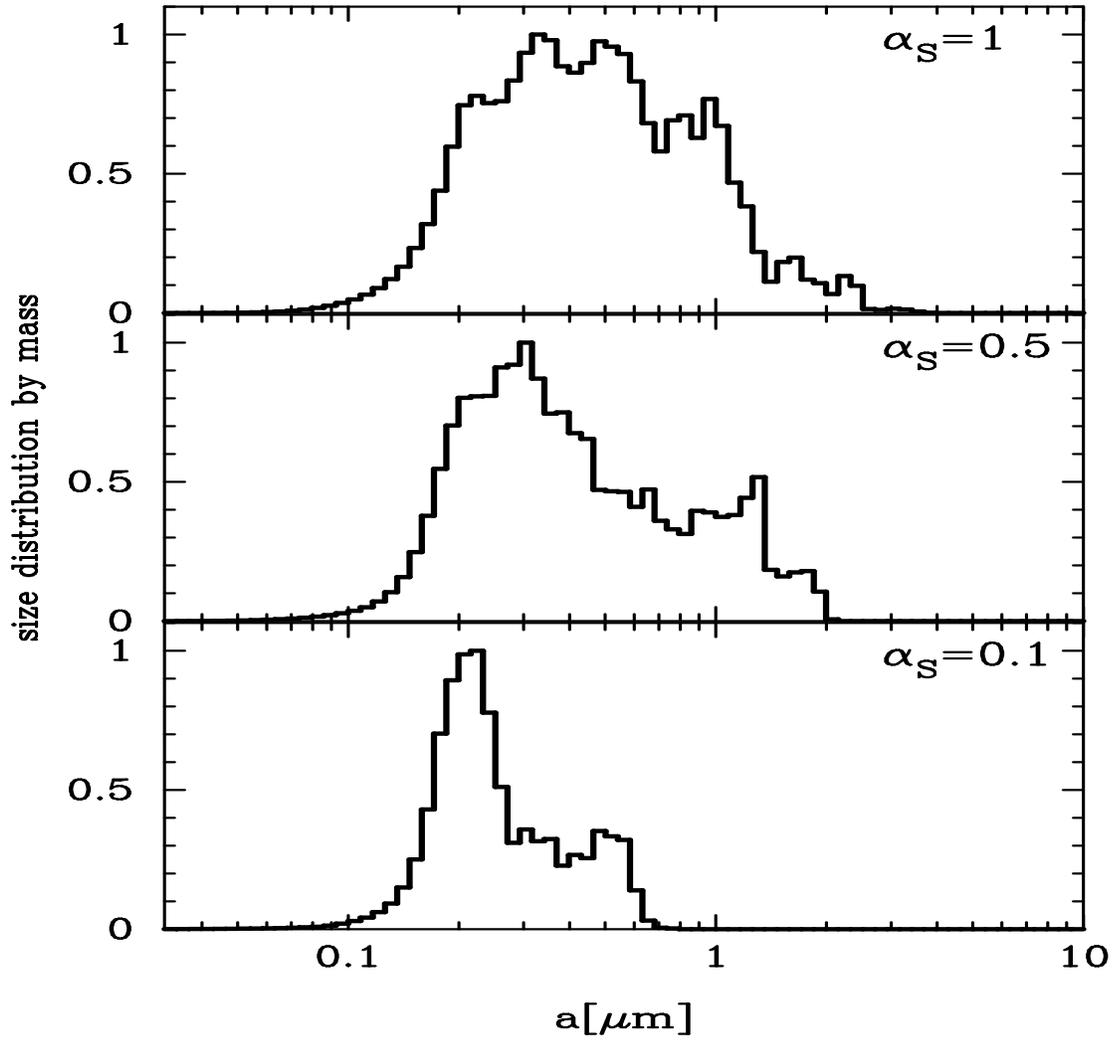}
\vspace{0.5cm}
\caption{Averaged size distributions of SiC grains by mass 
for given values of $\alpha_{{\rm S}}$,  
each of which is normalized by the maximum value.
\label{fig8}}
\end{figure}

\clearpage

\newpage

\begin{table}
\begin{center}
\caption{Reaction enthalpies $\Delta H{\arcdeg}$ for growth reactions of
 SiC cluster with size {\it n}=2 to 3. \label{table1}}
\vspace{0.5cm}
\begin{tabular}{|l|rr|}
\tableline
 & \multicolumn{2}{|c|}{$\Delta H{\arcdeg}$ [kJ mol$^{-1}$]} \\
\multicolumn{1}{|c|}{Reaction} & \multicolumn{1}{|l}{1000K} & \multicolumn{1}{l|}{1500K}  \\
\tableline
$\mathrm{SiC}+\mathrm{SiC} \rightarrow \mathrm{Si_{2}C_{2}}$ & -751.5 & -751.0 \\
$\mathrm{SiC}+\mathrm{Si_{2}C_{2}} \rightarrow \mathrm{Si_{3}C_{3}}$ & -518.8 & -514.5 \\
$\mathrm{SiC_{2}}+\mathrm{SiC} \rightarrow \mathrm{Si_{2}C_{2}}+\mathrm{C}$ & 76.98  & 78.12 \\
$\mathrm{SiC_{2}}+\mathrm{Si_{2}C_{2}} \rightarrow \mathrm{Si_{3}C_{3}}+\mathrm{C}$ & 309.6  & 314.7 \\
$\mathrm{SiC_{3}}+\mathrm{SiC} \rightarrow \mathrm{Si_{2}C_{2}}+\mathrm{C_{2}}$ & -54.88 & -58.76 \\
$\mathrm{SiC_{3}}+\mathrm{Si_{2}C_{2}} \rightarrow \mathrm{Si_{3}C_{3}}+\mathrm{C_{2}}$ & 177.8 & 177.8 \\
$\mathrm{SiC_{4}}+\mathrm{SiC} \rightarrow \mathrm{Si_{2}C_{2}}+\mathrm{C_{3}}$ & -123.0 & -131.1 \\
$\mathrm{SiC_{4}}+\mathrm{Si_{2}C_{2}} \rightarrow \mathrm{Si_{3}C_{3}}+\mathrm{C_{3}}$ & 109.7 & 105.4 \\
$\mathrm{Si_{2}C}+\mathrm{SiC} \rightarrow \mathrm{Si_{2}C_{2}}+\mathrm{Si}$ & -110.8 & -109.8 \\
$\mathrm{Si_{2}C}+\mathrm{Si_{2}C_{2}} \rightarrow \mathrm{Si_{3}C_{3}}+\mathrm{Si}$ & 121.9 & 126.7 \\
$\mathrm{Si_{2}C_{2}}+\mathrm{Si_{2}C_{2}} \rightarrow \mathrm{Si_{3}C_{3}}+\mathrm{SiC}$ & 232.7 & 236.5 \\
$\mathrm{Si_{2}C_{3}}+\mathrm{SiC} \rightarrow \mathrm{Si_{2}C_{2}}+\mathrm{SiC_{2}}$ & -199.5 & -202.6 \\
$\mathrm{Si_{2}C_{3}}+\mathrm{SiC} \rightarrow \mathrm{Si_{3}C_{3}}+\mathrm{C}$ & 110.1 & 112.1 \\
$\mathrm{Si_{2}C_{3}}+\mathrm{Si_{2}C_{2}} \rightarrow \mathrm{Si_{3}C_{3}}+\mathrm{SiC_{2}}$ & 33.14 & 33.97 \\
$\mathrm{Si_{3}C}+\mathrm{SiC} \rightarrow \mathrm{Si_{2}C_{2}}+\mathrm{Si_{2}}$ & -53.73 & -52.46 \\
$\mathrm{Si_{3}C}+\mathrm{Si_{2}C_{2}} \rightarrow \mathrm{Si_{3}C_{3}}+\mathrm{Si_{2}}$ & 178.9 & 184.1 \\
$\mathrm{Si_{3}C_{2}}+\mathrm{SiC} \rightarrow \mathrm{Si_{3}C_{3}}+\mathrm{Si}$ & -122.4 & -119.9 \\
$\mathrm{Si_{4}C}+\mathrm{SiC} \rightarrow \mathrm{Si_{2}C_{2}}+\mathrm{Si_{3}}$ & -122.6 & -124.9 \\
$\mathrm{Si_{4}C}+\mathrm{Si_{2}C_{2}} \rightarrow \mathrm{Si_{3}C_{3}}+\mathrm{Si_{3}}$ & 110.1 & 111.6 \\
$\mathrm{Si_{4}C_{2}}+\mathrm{SiC} \rightarrow \mathrm{Si_{2}C_{2}}+\mathrm{Si_{3}C}$ & -195.5 & -199.5 \\
$\mathrm{Si_{4}C_{2}}+\mathrm{SiC} \rightarrow \mathrm{Si_{3}C_{3}}+\mathrm{Si_{2}}$ & -16.53 & -15.41 \\
$\mathrm{Si_{4}C_{2}}+\mathrm{Si_{2}C_{2}} \rightarrow \mathrm{Si_{3}C_{3}}+\mathrm{Si_{3}C}$ & 37.19 & 37.04 \\
$\mathrm{Si_{5}C}+\mathrm{SiC} \rightarrow \mathrm{Si_{2}C_{2}}+\mathrm{Si}+\mathrm{Si_{3}}$ & 352.2 & 347.9 \\
$\mathrm{Si_{5}C}+\mathrm{Si_{2}C_{2}} \rightarrow \mathrm{Si_{3}C_{3}}+\mathrm{Si}+\mathrm{Si_{3}}$ & 584.8 & 584.5 \\
\tableline
\end{tabular}
\end{center}
\end{table}

\clearpage

\newpage

\begin{table}
\begin{center}
\caption{Reaction enthalpies $\Delta H{\arcdeg}$ for growth reactions of solid SiC.
\label{table2}}
\vspace{0.5cm}
\begin{tabular}{|l|rr|}
\tableline
 & \multicolumn{2}{|c|}{$\Delta H{\arcdeg}$ [kJ mol$^{-1}$]} \\
\multicolumn{1}{|c|}{Reaction} & \multicolumn{1}{|l}{1000K} & \multicolumn{1}{l|}{1500K} \\
\tableline
$\mathrm{SiC}+\mathrm{SiC(s)} \rightarrow 2\mathrm{SiC(s)}$ & -793.8 & -788.5 \\
$\mathrm{SiC_{2}}+\mathrm{SiC(s)} \rightarrow 2\mathrm{SiC(s)}+\mathrm{C}$ & 34.64 & 40.66 \\
$\mathrm{SiC_{3}}+\mathrm{SiC(s)} \rightarrow 2\mathrm{SiC(s)}+\mathrm{C_{2}}$ & -97.22 & -96.22 \\
$\mathrm{SiC_{4}}+\mathrm{SiC(s)} \rightarrow 2\mathrm{SiC(s)}+\mathrm{C_{3}}$ & -165.3 & -168.6 \\
$\mathrm{Si_{2}C}+\mathrm{SiC(s)} \rightarrow 2\mathrm{SiC(s)}+\mathrm{Si}$ & -153.1 & -147.3 \\
$\mathrm{Si_{2}C_{2}}+\mathrm{SiC(s)} \rightarrow 3\mathrm{SiC(s)}$ & -836.2 & -826.0 \\
$\mathrm{Si_{2}C_{2}}+\mathrm{SiC(s)} \rightarrow 2\mathrm{SiC(s)}+\mathrm{SiC}$ & -42.34 & -37.47 \\
$\mathrm{Si_{2}C_{3}}+\mathrm{SiC(s)} \rightarrow 3\mathrm{SiC(s)}+\mathrm{C}$ & -207.2 & -199.4 \\
$\mathrm{Si_{2}C_{3}}+\mathrm{SiC(s)} \rightarrow 2\mathrm{SiC(s)}+\mathrm{SiC_{2}}$ & -241.9 & -240.0 \\
$\mathrm{Si_{3}C}+\mathrm{SiC(s)} \rightarrow 2\mathrm{SiC(s)}+\mathrm{Si_{2}}$ & -96.07 & -89.92 \\
$\mathrm{Si_{3}C_{2}}+\mathrm{SiC(s)} \rightarrow 2\mathrm{SiC(s)}+\mathrm{Si_{2}C}$ & -286.6 & -284.1 \\
$\mathrm{Si_{3}C_{2}}+\mathrm{SiC(s)} \rightarrow 3\mathrm{SiC(s)}+\mathrm{Si}$ & -439.8 & -431.3 \\
$\mathrm{Si_{4}C}+\mathrm{SiC(s)} \rightarrow 2\mathrm{SiC(s)}+\mathrm{Si_{3}}$ & -164.9 & -162.4 \\
$\mathrm{Si_{4}C_{2}}+\mathrm{SiC(s)} \rightarrow 2\mathrm{SiC(s)}+\mathrm{Si_{3}C}$ & -237.8 & -237.0 \\
$\mathrm{Si_{4}C_{2}}+\mathrm{SiC(s)} \rightarrow 3\mathrm{SiC(s)}+\mathrm{Si_{2}}$ & -333.9 & -326.9 \\
$\mathrm{Si_{5}C}+\mathrm{SiC(s)} \rightarrow 2\mathrm{SiC(s)}+\mathrm{Si}+\mathrm{Si_{3}}$ & 309.8 & 310.4 \\
\tableline
\end{tabular}
\end{center}
Notes. SiC(s) denotes solid SiC which is assmued to be $\beta$-SiC.
\end{table}

\vspace{4cm}

\begin{table}
\begin{center}
\caption{Coefficients for calculating  
$\Delta G^{0}_{i}/k_{\rm B}T_{\rm v}$ in equation (5).
\label{table3}}
\begin{tabular}{ccccccc}
\tableline\tableline
i & $\alpha_{i}$ & $\beta_{i}$ & $\gamma_{i}$ & $\delta_{i}$ & $\eta_{i}$ & $\epsilon_{i}$ \\
\tableline
1 & 3.11656$\times$10$^{1}$ & -1.12418$\times$10$^{0}$ & 1.71784$\times$10$^{2}$ & -1.01082$\times$10$^{5}$ & 1.01432$\times$10$^{6}$ & -8.78548$\times$10$^{7}$ \\
2 & -2.96847$\times$10$^{3}$ & 2.79219$\times$10$^{2}$ & 5.55156$\times$10$^{4}$ & -8.91964$\times$10$^{5}$ & 1.15759$\times$10$^{8}$ & -1.28618$\times$10$^{10}$ \\
\tableline
\end{tabular}
\end{center}
\end{table}



\end{document}